\documentclass[twocolumn]{bmcart}

\usepackage{amsthm,amsmath}
\RequirePackage{natbib}
\RequirePackage{hyperref}
\usepackage[utf8]{inputenc} 
\usepackage{epsfig,amsmath}

\startlocaldefs

\newcommand{\be}{\begin{equation}}
\newcommand{\ee}{\end{equation}}
\newcommand{\ba}{\begin{eqnarray}}
\newcommand{\ea}{\end{eqnarray}}
\newcommand{\bal}{\begin{align}}
\newcommand{\eal}{\end{align}}

\newcommand{\bi}{\begin{itemize}}
\newcommand{\ei}{\end{itemize}}
\newcommand{\bfi}{\begin{figure}}
\newcommand{\efi}{\end{figure}}
\newcommand{\bn}{\begin{enumerate}}
\newcommand{\en}{\end{enumerate}}

\newcommand{\red}[1]{\textcolor{black}{#1}}
\newcommand{\blue}[1]{\textcolor{black}{#1}}

\endlocaldefs

\begin{document}

\begin{frontmatter}

\begin{fmbox}
\dochead{Methodology Article}


\title{Bayesian uncertainty analysis for complex systems biology models: emulation, global parameter searches and evaluation of gene functions.}


\author[
   addressref={aff1},                   
   corref={aff1},                       
   email={i.r.vernon@durham.co.uk}   
]{\inits{I}\fnm{Ian} \snm{Vernon}}
\author[
   addressref={aff2},
   noteref={n1},                        
   email={junli.liu@durham.ac.uk}
]{\inits{J}\fnm{Junli} \snm{Liu}}
\author[
   addressref={aff1},
   email={michael.goldstein@durham.ac.uk}
]{\inits{M}\fnm{Michael} \snm{Goldstein}}
\author[
   addressref={aff3},
   email={J.H.Rowe@sheffield.ac.uk }
]{\inits{J}\fnm{James} \snm{Rowe}}
\author[
   addressref={aff2},
   email={j.f.topping@durham.ac.uk}
]{\inits{J}\fnm{Jen} \snm{Topping}}
\author[
   addressref={aff2},
         noteref={n5},                        
   email={keith.lindsey@durham.ac.uk}
]{\inits{K}\fnm{Keith} \snm{Lindsey}}


\address[id=aff1]{
  \orgname{Department of Mathematical Sciences, Durham University}, 
  \street{South Road},
  \postcode{DH1 3LE},
  \city{Durham},
  \cny{UK}
}
\address[id=aff2]{%
  \orgname{Department of Biosciences, Durham University},
  \street{South Road},
  \postcode{DH1 3LE},
  \city{Durham},
  \cny{UK}
}


\address[id=aff3]{%
  \orgname{Department of Biosciences, Durham University, South Road, Durham, DH1 3LE, UK, current address: Department of Molecular Biology and Biotechnology, University of Sheffield, Firth Court},
  \street{Western Bank},
  \postcode{S10 2TN},
  \city{Sheffield},
  \cny{UK}
}


\begin{artnotes}
\note[id=n1]{Joint corresponding author: \href{mailto:junli.liu@durham.ac.uk}{junli.liu@durham.ac.uk}, {          } } 
\note[id=n5]{\href{mailto:michael.goldstein@durham.ac.uk}{michael.goldstein@durham.ac.uk}, \href{mailto:J.H.Rowe@sheffield.ac.uk }{J.H.Rowe@sheffield.ac.uk},
\newline \href{mailto:j.f.topping@durham.ac.uk}{j.f.topping@durham.ac.uk}, \href{mailto:keith.lindsey@durham.ac.uk}{keith.lindsey@durham.ac.uk}.} 
\end{artnotes}



\begin{abstractbox}

\begin{abstract} 
\parttitle{Background} 
Many mathematical models have now been employed across every area of systems biology.
These models increasingly involve large numbers of unknown parameters, have complex structure which can result in substantial evaluation time relative to the needs of the analysis, and need to be compared to observed data of various forms.
The correct analysis of such models usually requires a global parameter search, over a 
high dimensional parameter space, that incorporates and respects the most important sources of uncertainty. 
This can be an extremely difficult task, but it is essential for any meaningful inference or prediction to be made about any biological system. It hence represents a fundamental challenge for the whole of systems biology.

\parttitle{Results} 
Bayesian statistical methodology for the uncertainty analysis of complex models is introduced, which is designed to address the high dimensional global parameter search problem.
Bayesian emulators that mimic the systems biology model but which are extremely fast to evaluate are embeded within an 
iterative history match: an efficient method to search high dimensional spaces within a more formal statistical setting, 
while incorporating major sources of uncertainty. 
The approach is demonstrated via application to a model of hormonal crosstalk in Arabidopsis root development, which has 32 rate parameters, 
for which we identify the sets of rate parameter values that lead to acceptable matches between model output and observed trend data. 
The multiple insights into the model's structure that this analysis provides are discussed. The methodology is applied to a second related model, and the biological consequences of the resulting comparison, including the evaluation of gene functions, are described.

\parttitle{Conclusions} 
Bayesian uncertainty analysis for complex models using both emulators and history matching is shown to be a powerful technique that can greatly 
aid the study of a large class of systems biology models. 
It both provides insight into model behaviour and identifies the sets of rate parameters of interest.

\end{abstract}




\end{abstractbox}
\end{fmbox}

\end{frontmatter}



\section*{Background}

\subsection*{Fundamental challenges facing systems biology}

Recent advances in genome sequencing techniques, a variety of `omic' techniques and bioinformatic analyses, have led to an explosion of systems-wide biological data. Thus, identification of molecular components at the genome scale based on biological data has become possible. However, a major challenge in biology is to analyse and predict how functions in cells emerge from interactions between molecular components. Computational and mathematical modelling provide compelling tools to study the nonlinear dynamics of these complex interactions \cite{sb_philo:2007aa}. A particular example is kinetic modelling, in which the kinetics of each biological reaction are described in accordance with the corresponding biological process, and the properties of the whole system are described using differential equations: a common tool for analysing biological systems \cite{Alves:2006aa, Jamshidi:2008aa, Smallbone:2010aa}. 

A critical problem found in the mathematical modelling of many complex biological systems, which is of particular severity in kinetic modelling, 
is that the models often contain large numbers of uncertain parameters (a common type being reaction rate parameters).
In most cases, such kinetic parameters cannot be directly measured as experiments typically measure concentrations rather than rates. Even when such parameters can be measured `directly', this is usually in experimental conditions that are significantly different from the 
cellular environment we wish to study. Therefore, we have to compare the mathematical model's outputs with experimental observations, often in the form of measured concentrations and trends, and determine which values of the input or rate parameters will achieve an acceptable match between model and reality. This involves consideration of several sources of uncertainty including observation error, biological variability and the tolerance we place on the model's accuracy, known as the model discrepancy. It is vital that we perform a {\it global} parameter search for {\it all} input parameter settings that achieve an acceptable match. This is because a single solution for the rate parameter values may suggest certain biological implications and give particular predictions for future experiments, both of which could be gravely misleading were we to explore the parameter space further and find several alternative solutions that give radically different implications and predictions. This is a mistake that is disturbingly common. 

Unfortunately, performing global parameter searches over high dimensional spaces can be extremely challenging for several reasons, most notably: (a) the complex structure of the model and hence the complex way it imposes constraints on the parameters, (b) the substantial model evaluation time relative to the needs of the analysis, (c) the need for a careful assessment of an ``acceptable match" that incorporates appropriately all the complexities and uncertainties of the comparison between the model and the real system, and (d) high dimensional spaces, being extremely large, require {\it vast} numbers of model evaluations to explore. For example, some spatial models of root development \cite{NPH:NPH13421} require at least several minutes for a single evaluation. It is worth considering how large high dimensional spaces are: were we just to evaluate the model in question at the corners of the initial input space, in say 32 dimensions, we would require $2^{32} \simeq 4.3$ billion evaluations, which would take approximately 136 years if the model took 1 second per evaluation.
However, global parameter searches are essential for any meaningful inference or prediction to be made about the biological system. Therefore this represents a fundamental challenge for the whole of systems biology. This article describes practical methodology to address this problem, based on Bayesian statistics methodology for the uncertainty analysis of complex models~\cite{Vernon10_CS,galf_stat_sci,Craig97_Pressure,Kennedy01_Calibration}.

\subsection*{Bayesian emulation and uncertainty analysis}

The issues surrounding the analysis of complex models under uncertainty, and specifically the global parameter search problem, are
not unique to systems biology, and have been encountered in many different scientific disciplines. 
An area of Bayesian statistics has arisen to meet the demand of such analyses. 
This area, sometime referred to as the uncertainty analysis of computer models, centres around the construction of Bayesian emulators~\cite{Vernon10_CS,galf_stat_sci,Craig97_Pressure,Kennedy01_Calibration}.
An emulator is a statistical construct that mimics the scientific model in question, providing predictions of the model outputs with associated uncertainty, at as yet unevaluated input parameter settings. The emulator is however, extremely fast to evaluate~\cite{OHagan06_Tutorial}. It provides insight into the model's structure and, thanks to its speed, it can be used to help perform the global parameter search far more efficiently than approaches that just use the comparatively slow scientific model itself (for examples see \cite{Vernon10_CS,Craig97_Pressure,Yiannis_HIV_1,Williamson:2013aa,Higdon09_Coyote2,Oakley02_UncertainOutputs}).

Many analyses and corresponding parameter searches still fail because an appropriate measure of an acceptable match between model and reality is not defined. This can lead to the use of badly behaved objective functions that do not properly capture the desired match criteria, and which 
are often harder to explore in high dimensions, due to increased numbers of ridges, spikes and local minima. 
The Bayesian emulation methodology we introduce naturally incorporates more detailed statistical models of the difference between the model outputs and the observed data, which allow the inclusion of important sources of uncertainty such as observational error and model discrepancy, the later being the upfront acknowledgement of the limitations of the current model.
Various structures of increasing complexity are available for the representation of these uncertainties, depending on the requirements and importance of the study (see \cite{Vernon10_CS,Vernon10_CS_rej, asses_mod,Brynjarsdottir:2014aa, Goldstein09_Reify} for examples and discussion).

It is worth noting that due to their speed, the use of emulators would greatly improve the efficiency of many forms of analysis that a modeller may wish to perform, e.g. for a fully Bayesian MCMC analysis~\cite{Kennedy01_Calibration,Higdon04_prediction, Henderson:2009aa} or for more direct global 
parameter searches such as~\cite{Zamora-Sillero:2011aa}. 
However for high dimensional models, the particular strategy chosen for a parameter search is vital. Many approaches struggle 
due to being trapped in local minima (of which there may be many) or because they chase the scientifically spurious 
best match parameter setting. 
Here, we describe an efficient global parameter search method known as Bayesian history matching, which has proved very successful 
across a wide range of scientific disciplines including cosmology~\cite{Vernon10_CS,Vernon10_CS_rej,vernon_astro,
galf_stat_sci,2009:45:isipta,Vernon:2016aa}, epidemiology~\cite{Yiannis_HIV_1,Yiannis_HIV_2}, oil reservoir 
modelling~\cite{Craig96_Pressure,Craig97_Pressure,JAC_Handbook,JAC_sma_samp}, climate modelling~\cite{Williamson:2013aa}, environmental science~\cite{asses_mod} 
\blue{and traffic modelling~\cite{Boukouvalas:2014aa}}.

It utilises Bayesian emulators to reduce efficiently the input parameter space in iterations or waves, by identifying regions that are implausible as matches to the observed data,
with the objective of identifying  
all acceptable input parameter settings. It is a careful approach that avoids many of the traps of common parameter search techniques.

\subsection*{Hormonal crosstalk network in Arabidopsis root development}

Understanding how hormones and genes interact to coordinate plant growth is a major challenge in developmental biology. The activities of auxin, ethylene and cytokinin depend on the cellular context and exhibit either synergistic or antagonistic interactions. Previously, three of our authors developed a hormonal crosstalk network for a single Arabidopsis cell by iteratively combining modelling with experimental analysis~\cite{Liu10_crosstalk}. Kinetic modelling was used to analyse how such a network regulates auxin concentration in the Arabidopsis root, by controlling the relative contribution of auxin influx, biosynthesis and efflux; and by integrating auxin, ethylene and cytokinin signalling~\cite{Liu10_crosstalk}. Although some of the parameters in the model were based on experimental data, most parameters were chosen in an ad hoc way, by adjusting them to fit experimental data. Conditional on those somewhat ad hoc choices, it was shown that the hormonal crosstalk network quantitatively describes how the three hormones (auxin, ethylene, and cytokinin) interact via POLARIS peptide (PLS) \cite{A:2002aa,Chilley-PM:2006aa} to regulate plant root growth~\cite{Liu10_crosstalk}.

In this work we demonstrate the power of the Bayesian emulation methodology by applying it to the hormonal crosstalk network in Arabidopsis root development. Specifically, we explore the model's 32-dimensional parameter space, and identify the set of all acceptable matches between model outputs and experimental data, taking into account major sources of uncertainty. This provides much insight into the model's structure and the constraints imposed
on the rate parameters by the current set of observed data. We apply the methodology to a second, competing model, and hence are able to investigate gene functions robustly. As an example, our analysis suggest that, in the context of the hormonal crosstalk network, POLARIS peptide (PLS) must have a role in positively regulating auxin biosynthesis. 

The paper is organised as follows. In the Methods section we begin by defining a simple 1-dimensional toy model that we use to illustrate our definitions and to demonstrate the three main parts of the Bayesian methodology: linking the model to reality, Bayesian emulation, and history matching, \red{before going on to compare the strengths and weaknesses of Bayesian history matching to more standard approaches}. In the Results and Discussion section we describe in detail the application of this methodology to the full 32 dimensional Arabidopsis model, and discuss the relevant insights and biological implications obtained.

\section*{Methods}

\subsection*{Simple 1-dimensional exponential example}

Here we introduce a simple 1-dimensional exponential toy model example which we will use 
to illustrate our definitions of all the parts of a typical systems biology analysis, for example, the model, the input or rate parameters, observations with errors, model discrepancy, Bayesian emulators, implausibility measures and history matching.
Specifically, this 1-dimensional example will be used throughout this Methods section to demonstrate each of the three main parts of our approach: 
\smallskip
\bi
\item Linking the model to reality
\item Bayesian Emulation
\item History matching: a global parameter search
\ei
\smallskip

Say we are interested in the concentration of a chemical which evolves in time. We represent this concentration as 
$f_t(x)$ where $x$ is, for example, a reaction rate parameter and $t$ is time.
We model $f_t(x)$ with the differential equation:
\be\label{eq_fxt}
 \frac{df_t(x)}{dt} \;\;=\;\; x \; f_t(x)    
 \ee
 which in this case we can solve precisely to give
 \be\label{eq_fxt_an}
  f_t(x) \;\;=\;\; f_0 \; \exp{(x t)}  
\ee
We will temporarily assume the initial conditions are $f_0=f_{t=0}(x) = 1$.
The system runs from $t=0$ to $t=5$ and we are at first interested in the value of $f_t(x)$ at $t=3.5$.
This mathematical model features an input or rate parameter $x$, which we wish to learn about. We do this using a measurement of the real biological system at $t=3.5$ which we denote $z$, which corresponds to, but is not the same as, the model output $f_{t=3.5}(x)$.
Note that usually, for models of realistic complexity, we would not have the analytic solution for $f_t(x)$ given by equation~(\ref{eq_fxt_an}). Instead we would resort to a numerical integration method to solve equation~(\ref{eq_fxt}) that might require significant time for one model evaluation, ranging from less than a second to hours, days or even weeks, depending on the full complexity of the model~\cite{Vernon10_CS,Williamson:2013aa}.
Such a computational cost, for a single evaluation of the model, means that a full global parameter search is computationally infeasible, especially when the model has many rate parameters and therefore a high dimensional input space, which may require vast numbers of evaluations to explore fully.

\begin{figure*}
\vspace{-0.5cm}
\begin{center}
\hspace{-0.0cm}\includegraphics[scale=0.49,angle=0]{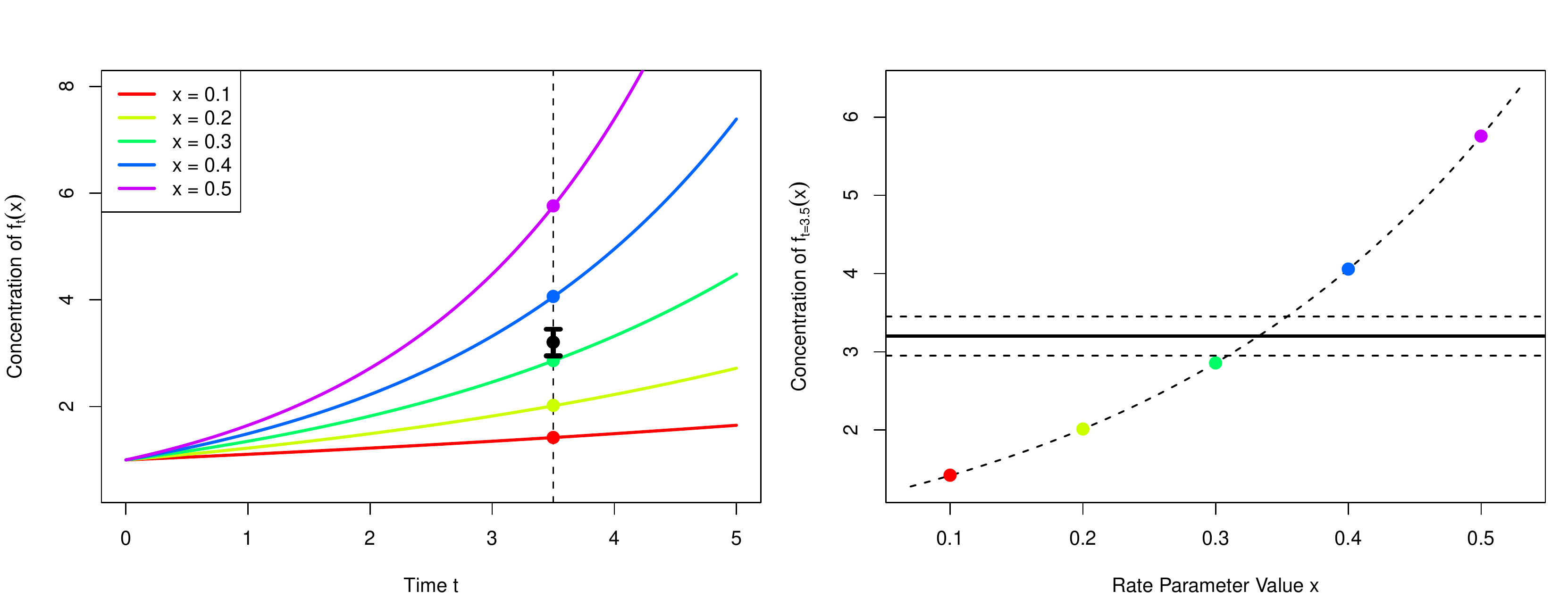} 
\caption{Left panel: five evaluations of the model $f_t(x)$ for different values of $x$ between $x = 0.1$ and $0.5$, coloured red to purple respectively, with time $t$ on the $x$-axis. The measurement of the real system at $t=3.5$ is denoted $z$, and is represented as the black point, 
with $\pm 3\sigma_e$ error bars representing observational error. Right panel: the model at $t=3.5$. The curved dashed line gives $f_{t=3.5}(x)$ but now it is represented purely as a function of the input parameter $x$ on the $x$-axis, 
with the red to purple points consistent with those in the left panel.
The observed data $z$ is given by the solid black horizontal line, with the $\pm 3\sigma_e$ error bars as the horizontal black dashed lines. We see that there will be acceptable values of $x$ approximately between 0.3 and 0.35.}\label{fig_1d_setup_mod}
\end{center}
\end{figure*}

We typically begin the analysis by exploring the model's behaviour for several different values of the unknown rate or input parameter $x$. 
Figure~\ref{fig_1d_setup_mod} (left panel), shows five evaluations of the model $f_t(x)$ for different values of $x$ between $x=0.1$ and $0.5$, coloured red to purple respectively, with time on the x-axis.
The measurement of the system is denoted $z$, and is represented as the black point in figure~\ref{fig_1d_setup_mod}, with $\pm3\sigma_e$ error bars representing observational error, defined precisely below. This measurement was made at $t=3.5$ shown as the vertical dashed line. 
The most important questions for the biologist at this point are: can the model match the observed data $z$ at all, and 
if so, what is the entire set of input parameter choices that give rise to acceptable matches between model output and observed data?
Figure~\ref{fig_1d_setup_mod} (right panel) represents this question as it now shows only $f_{t=3.5}(x)$ but now represented purely as a function of the input parameter $x$ on the x-axis, with the red to purple points consistent with those in the left panel.
The observed data $z$ is now represented as the solid black horizontal line, with the $\pm 3\sigma_e$ error bars as the horizontal black dashed lines. 
We see that there will be acceptable values of $x$ approximately between 0.3 and 0.35. 

For a general complex model $f_t(x)$, that possesses a large number of input or rate parameters and possibly several outputs, a full analysis of the model's behaviour encounters the following issues:
\smallskip
\bn
\item When comparing the model to observed data from the real biological system, an adequate statistical description of the link between model and reality, covering all major uncertainties, is required.
\item For complex models, the time taken to evaluate the model numerically is so long that an exhaustive exploration of the model's behaviour is not feasible.
\item The appropriate scientific goal should be to identify {\it all} locations in input parameter space that lead to acceptable fits between model and data, and not just find the location of a single good match. 
\en
\smallskip
Methods to address these three fundamental issues are described in the next three sections.

\subsection*{Model Discrepancy and Linking the model to reality}

Most systems biology models are developed to help explain and understand the behaviour of corresponding real world biological systems. 
An essential part of determining whether such models are adequate for this task is the process of comparing the model to experimental data.
As a comparison of this kind involves several uncertainties that cannot be ignored, it is therefore vital to develop a clearly defined statistical model for the link between systems biology model $f(x)$ and reality $z$.
This allows for a meaningful definition of an `acceptable' match between a model run and the observed data.
Here we describe a simple yet extremely useful statistical model for the link between the biological model and reality, that has been successfully applied in a variety of scientific disciplines, for example climate, cosmology, oil reservoir modelling and epidemiology~\cite{Williamson:2013aa,Vernon10_CS,Craig97_Pressure,Yiannis_HIV_1}.

The most recognisable source of uncertainty is that of observational or experimental error. 
We represent the uncertain quantities of interest in the real biological system as the vector $y$, which we will measure with a vector of errors $e$ to give the vector of observations $z$, such that
\be
z \;\;=\;\; y \; + \; e   \label{eq_zye}
\ee
where we represent the errors as additive, although more complex forms could be used. Note that $z$, $y$ and $e$ here represent vectors of random quantities, which will reduce to scalar random quantities if there is only one quantity of interest. A common specification \cite{Vernon10_CS} that we will employ here is to judge the errors to be independent from $y$, and unbiased with expectation ${\rm E}(e) =0$ and, for the scalar case, ${\rm Var}(e) = \sigma_e^2$.

An important distinction to make is between the model of the biological system, represented as the vector $f(x)$, and the system itself $y$. We represent the difference between 
these using a {\it model discrepancy} term as follows. Even were we to evaluate the model $f(x)$ at its best possible choice of input $x^*$, the output $f(x^*)$ would still not be in agreement with the real biological system value $y$, due to the many simplifications and approximations of the model. Hence we state that:
\be
y \;\;=\;\; f(x^*) \; + \; \epsilon     \label{eq_yfep}
\ee
where the $\epsilon$ is the model discrepancy: a vector of uncertain quantities that represents directly the difference between the model and the real system. 
Again we treat $y$, $f$, $x^*$ and $\epsilon$ as vectors of random quantities.
A simple and popular specification \cite{Vernon10_CS} would be to judge that $\epsilon$ is independent of $f(x^*)$, $x^*$ and $e$, with ${\rm E}(\epsilon) =0$ and, in the scalar case, ${\rm Var}(\epsilon) = \sigma_{\epsilon}^2$. In a multivariate setting, where $f(x)$ describes a vector of outputs (for example, with each output labelled by time $t$), the vector $\epsilon$ may have an intricate structure, possessing non-zero covariances between components of $\epsilon$. This could capture the related deficiencies of the model across differing time points. Various structures of increasing complexity are available (for examples see \cite{Craig97_Pressure,Kennedy01_Calibration,Vernon10_CS}), along with 
methods for specification of their components~\cite{Vernon10_CS,asses_mod}. 

While the explicit inclusion of the model discrepancy term $\epsilon$ is unfamiliar, it is now standard practice in the statistical literature for complex models \cite{Craig01_Forecasting,Kennedy01_Calibration,galf_stat_sci,Brynjarsdottir:2014aa}. Furthermore, any analysis performed without such a term \red{is implicitly conditioned with the statement} ``given the model is a perfect representation of reality for some value of the inputs $x$", a statement that is rarely true. 
\red{The model discrepancy allows us to perform a richer analysis than before as we can now include any extra knowledge we have about the model's deficiencies to improve our modelling of reality $y$, through the joint structure of $\epsilon$ (see for example \cite{asses_mod,Vernon10_CS}). This is especially important in prediction: as a simple example, if our model undershot every auxin output we have measured so far, we may suspect that it will undershoot future measurements of auxin also, and may wish to build this into our prediction for future $y$ \cite{Craig01_Forecasting}.}

We can specify probabilistic attributes of $\epsilon$ a priori, or learn about them by comparing to observed data. For direct specification, there are often various simple experiments that can be performed on the model itself to obtain assessments of $\sigma_{\epsilon}$ and other aspects if necessary. For example, often models are run from exact initial conditions, so performing a set of exploratory model evaluations with the initial conditions appropriately perturbed would provide a lower bound on $\sigma_{\epsilon}$. See~\cite{asses_mod} where several such assessment methods are demonstrated, for more details.

\subsubsection*{1-dimensional example}

\begin{figure*}
\vspace{-0.5cm}
\begin{center}
\hspace{-0.0cm}\includegraphics[scale=0.49,angle=0]{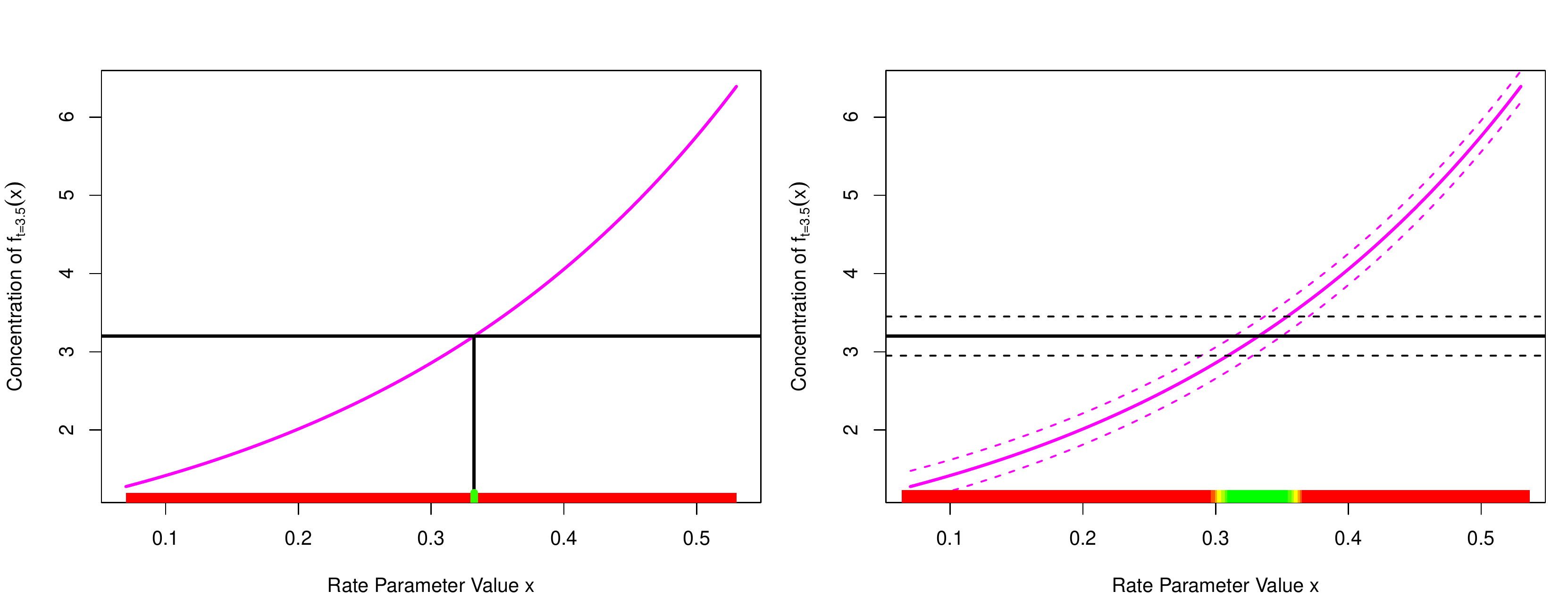} 
\caption{Left panel: the unrealistic case when both the observation error $e$ and model discrepancy $\epsilon$ are ignored. The model $f(x)$ is given by the purple line, the observed data $z$ by the horizontal black line. Here only one value of $x$ can be viewed as acceptable (coloured green) while all others are unacceptable (red), however a repeat of the experiment would yield a slightly different $z$ and hence $x$. Right panel: the more realistic situation where we include both observation error $e$ (the black dashed lines represent $z\pm 3 \sigma_e$) and model discrepancy $\epsilon$ (the purple dashed lines show $f(x) \pm 3\sigma_{\epsilon}$). There is now a range of acceptable values for $x$ (green points) with borderline/unacceptable points in yellow/red. It now may be possible to simultaneously match multiple outputs.}\label{fig_1d_mod_dis}
\end{center}
\end{figure*}

Figure~\ref{fig_1d_mod_dis} (left panel) shows the case for the simple 1-dimensional exponential example model when both the observation error $e$ and model discrepancy $\epsilon$ are ignored. The model $f(x)$ is given by the purple line, while the observed data $z$ is given by the horizontal black line. 
Here only one value of $x$ can be viewed as acceptable (coloured green) while all others are unacceptable (red). This particular value of $x$ is not unique in that if we were to perform the measurement again, due to measurement error we would get a different value for $z$ and hence for $x$. More importantly, if the model had a second output, say corresponding to a different time, that also depended on the same input $x$, we would be extremely unlikely to be able to match both outputs to their measurements as we would have to obtain exact matches simultaneously for precisely the same value of $x$. Inferences and predictions about the biological system made from this case, using this value of $x$ are not trustworthy.

Figure~\ref{fig_1d_mod_dis} (right panel) shows the far more realistic situation where we include both observation error $e$ (the black dashed lines represent $z\pm 3 \sigma_e$) and model discrepancy $\epsilon$ (the purple dashed lines show $f(x) \pm 3\sigma_{\epsilon}$). As we have taken into account both major types of uncertainty there is now a range of acceptable values for $x$ (green points) with borderline/unacceptable points in yellow/red. If we were to consider additional outputs of the model, we still have a chance to match them simultaneously to data for a subset of the currently acceptable points. If on the other hand we cannot find any acceptable points $x$ even given the uncertainties represented by $e$ and $\epsilon$, then we can state that the model is inconsistent with the observed data and therefore most likely based on incorrect biological principles. 
Further, inclusion of the observation error and model discrepancy terms often aids a global parameter search as they tend to smooth the likelihood surface (or comparable objective function), making it both easier to explore while simultaneously more robust. They also help reduce the temptation to chase the \red{often scientifically misleading} global minimum, such as the lone green point in figure~\ref{fig_1d_mod_dis} (left panel), instead suggesting that the identification of a set of acceptable input points is the appropriate goal for such a search (see the green points in figure~\ref{fig_1d_mod_dis} (right panel)).

\subsection*{Bayesian emulation of systems biology models}

Many complex mathematical models have been developed and employed within the area of systems biology.
Often these models have high dimensional input spaces in that they posses several input parameters, for example reaction rate parameters, that must be specified in order to evaluate the model. We represent the list of such inputs as the vector $x$, with individual inputs as $x_k$ with $k=1,\dots ,d$.
The model may have any number of outputs, denoted as the vector $f(x)$, with individual outputs as $f_i(x)$ with $i=1,\dots ,q$, the behaviour of which we want to investigate, possibly comparing some of these to observed data. For example, the index $i$ may label the different times we are interested in, or the different chemical outputs of the model, or both.
Most models are complex enough that they require numerical integration methods to solve, and hence take appreciable time to evaluate. 
This evaluation time can range anywhere from less than a second to minutes, hours or even days for highly sophisticated models: our approach is applicable in any of these cases, and adds more value as the dimensionality and evaluation time of the model increases.

A Bayesian emulator is a fast, approximate mimic of the full systems biology model. It gives insight into the structure of the model's behaviour and can be used instead of the model in many complex calculations. The emulator gives a prediction of what the model's output $f(x)$ will be at a yet to be evaluated input point $x$, and additionally provides an associated uncertainty for that prediction (these are often expressed as posterior distributions, or simply expectations and variances in some cases). 
Critically an emulator is extremely fast to evaluate as it only requires a few matrix multiplications, and hence can be used to explore the input space more fully, as for example in a global parameter search.

A popular choice for the Bayesian emulator for model $f(x)$, which has individual outputs $f_i(x)$, $i=1\dots q$, is structured as follows:
\be
\label{eq_emulator}
f_i(x) = \sum_j \beta_{ij}  g_{ij}(x_{A_i}) + u_i(x_{A_i}) + w_i(x) 
\ee
where the active variables $x_{A_i}$ are a subset of the inputs $x$ that are most influential for output $f_i(x)$. The first term on the right hand side of 
the emulator equation~(\ref{eq_emulator}) is a regression term, where 
$g_{ij}$ are known deterministic 
functions of $x_{A_i}$, a common choice being low order polynomials, and $\beta_{ij}$ are unknown scalar regression coefficients. The second term, $u_i(x_{A_i})$ is a Gaussian process over $x_{A_i}$ (or in a less fully specified version, a weakly second order stationary stochastic process), which means that if we choose a finite set of inputs $\{x^{(1)}_{A_i},\dots,x^{(s)}_{A_i} \}$, the uncertain outputs $u_i(x^{(1)}_{A_i}),\dots,u_i(x^{(s)}_{A_i})$ will have a multivariate normal distribution with a covariance matrix constructed from an appropriately chosen covariance function, for example:
\be
\label{eq_corr} 
\! \!{\rm Cov}(u_i(x_{A_i}),u_i(x'_{A_i})) = \sigma^2_{u_i} {\rm exp}\left\{- \frac{\|x_{A_i}-x_{A_i}'\|^2}{  \theta_i^2}\right\}  \!
\ee
where $\sigma^2_{u_i}$ and $\theta_i$ are the variance and correlation length of $u_i(x_{A_i})$ which must be specified \cite{Vernon10_CS}. 
The third term $w_i(x)$ is a nugget, a white noise process uncorrelated with $\beta_{ij}$, $u_i(x_{A_i})$ and itself such that
\be
\label{eq_nugget} 
\! \!{\rm Cov}(w_i(x),w_i(x')) = \left\{  \begin{array}{cl} 
						\sigma^2_{w_i} & {\rm if} \quad x=x' \\
						0 & {\rm otherwise}
						\end{array}
						\right.
\ee
with expectation zero and ${\rm Var} (w_i(x)) =  \sigma^2_{w_i}$, that represents the effects of the remaining inactive input variables~\cite{Vernon10_CS}.

The emulator, as given by equation~(\ref{eq_emulator}), possesses various desirable features. The regression term, given by $\sum_j \beta_{ij}  g_{ij}(x_{A_i})$, is often chosen to represent say a third order polynomial in the active inputs. This would attempt to mimic the large scale global behaviour of the function $f_i(x)$, and in many cases, will capture a large proportion of the model's structure. (It is worth noting that reasonably accurate emulators can 
often be constructed just using regression models, for example using the lm() function in R. This can be a sensible first step, before one attempts the construction
of a full emulator of the form given in equation~(\ref{eq_emulator}).)
The second term $u_i(x_{A_i})$, the Gaussian process, mimics the local behaviour of $f_i(x)$ and specifically its local deviations from the third order polynomial given by the regression terms. We can choose the list of active inputs $x_{A_i}$ using various statistical techniques for example, classical linear model fitting criteria such as AIC or BIC \citep{Vernon10_CS}. A list of say $p$ active inputs for a particular output 
$f_i(x)$ means that we have reduced the input dimension from $d$ to $p$ dimensions, which can result in large efficiency gains.
The small remaining effect of the inactive inputs is captured by the third term $w_i(x)$ in equation~(\ref{eq_emulator}).

\begin{figure*}
\vspace{-0.5cm}
\begin{center}
\hspace{-0.0cm}\includegraphics[scale=0.52,angle=0]{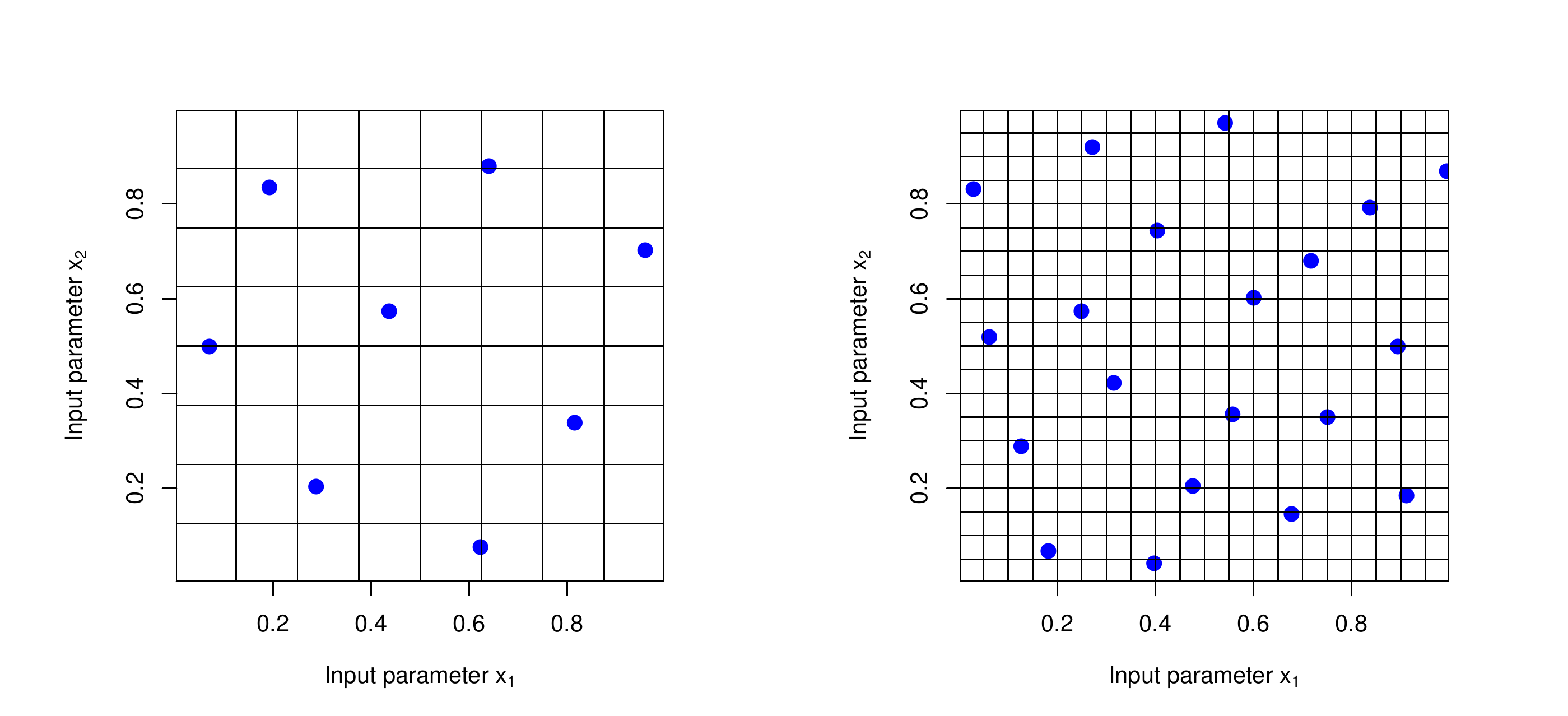} 
\caption{Maximum minimum distance Latin hypercube designs of size $n=8$ (left panel) and $n=20$ (right panel). The blue points represent 
locations in rate or input parameter space where we would run the full systems biology model. These designs are both space filling and
approximately orthogonal, both desirable features for fitting emulators. Note that the construction process of a Latin hypercube ensures that there is a blue point within each of the $n$ subintervals 
of both inputs, ensuring excellent coverage.}\label{fig_lhc}
\end{center}
\end{figure*}

We proceed by performing an initial set of carefully chosen model evaluations, often picked to be `space filling', i.e. well spread out over the input space. For example we may use a maximin \red{Latin} hypercube design, an approximately orthogonal design which attempts to ensure there are no large holes in-between the run locations (see figure~\ref{fig_lhc} and \cite{SWMW89_DACE,Santner03_DACE,Currin91_BayesDACE}). An $n$ point \red{Latin} hypercube design is created by dividing the range of each input into $n$ sub-intervals, and placing points to ensure there is only ever one point in each sub-interval (this can be done using the lhs() function in R~\citep{R-Core-Team:2015aa}).
Many such \red{Latin} hypercube designs are generated and the one with maximum minimum distance between points is chosen. 

We then fit the emulator given by equation~(\ref{eq_emulator}) to the set of model runs using our favourite statistical tools, guided by expert judgement. Specifically we would prefer a fully Bayesian approach if we required full probability distributions~\cite{Kennedy01_Calibration}, and a Bayes Linear approach~\cite{Goldstein_99,Goldstein07_BayesLinearBook}, which we will describe below, if we required purely expectations, variances and covariances of $f(x)$. 
We make certain pragmatic choices in the emulator construction process, for example, while we keep the regression coefficients $\beta_{ij}$ uncertain, we may directly specify $\sigma^2_{u_i}$, $\sigma^2_{w_i}$ and 
$\theta_i$ a priori, or use suitable plugin estimates~\cite{Vernon10_CS}.

The emulators then provide an expectation and variance for the value of $f(x)$ at an unexplored input point $x$. We can test the emulators using a series of diagnostics, for example checking their prediction accuracy over a new batch of runs~\cite{Tony_EmDiag}. See~\cite{OHagan06_Tutorial} for an introduction and \cite{galf_stat_sci,Vernon10_CS,Kennedy01_Calibration} for detailed descriptions of emulator construction.

While there are several approaches to emulator construction, our preferred choice is to use Bayes Linear methods, which is a more tractable version of Bayesian statistics which requires a far simpler prior specification and analysis~\cite{Goldstein_99,Goldstein07_BayesLinearBook}. It deals purely with expectations, variances and covariances of all uncertain quantities, and uses the following update equations to adjust our beliefs in the light of new data. Say we had performed an initial wave of $n$ runs at input locations $x^{(1)}, x^{(2)},\dots,x^{(n)}$ giving a column vector of model output values $D_i = (f_i(x^{(1)}), f_i(x^{(2)}),\dots,f_i(x^{(n)}))^T$, where $i$ labels the model output. We obtain the adjusted expectation and variance for $f_i(x)$ at new input point $x$ using:
\begin{align}
& \hspace{-0.75cm}{\rm E}_{D_i}(f_i(x)) = \label{eq_BLE}\\
&\hspace{-0.5cm}{\rm E}(f_i(x)) + {\rm Cov}( f_i(x), D_i) {\rm Var}(D_i)^{-1} (D_i - {\rm E}(D_i)) \nonumber \\
&\hspace{-0.75cm}{\rm Var}_{D_i}(f_i(x)) =  \label{eq_BLV} \\
&\hspace{-0.5cm}{\rm Var}(f_i(x)) - {\rm Cov}( f_i(x), D_i) {\rm Var}(D_i)^{-1} {\rm Cov}( D_i,f_i(x)) \nonumber \\ \nonumber
\end{align}
All quantities on the right hand side of equations~(\ref{eq_BLE}) and (\ref{eq_BLV}) can be calculated from equations~(\ref{eq_emulator}) and (\ref{eq_corr})
combined with prior specifications for ${\rm E}(\beta_{ij})$, ${\rm Var}(\beta_{ij})$, $\sigma^2_{u_i}$, $\sigma^2_{w_i}$ and $\theta_i$. Note that we could have used the entire collection of model outputs $D =\{D_1,D_2, \dots , D_q\}$ instead of just $D_i$ in equations~(\ref{eq_BLE}) and (\ref{eq_BLV}), if we had specified a more complex, multivariate emulator~\cite{Rougier:2008aa}. 

${\rm E}_{D_i}(f_i(x))$ and ${\rm Var}_{D_i}(f_i(x))$ are used directly in the implausibility measures used for the global parameter searches described below.

\subsubsection*{1-dimensional example}

\begin{figure*}
\vspace{-0.5cm}
\begin{center}
\hspace{-0.0cm}\includegraphics[scale=0.49,angle=0]{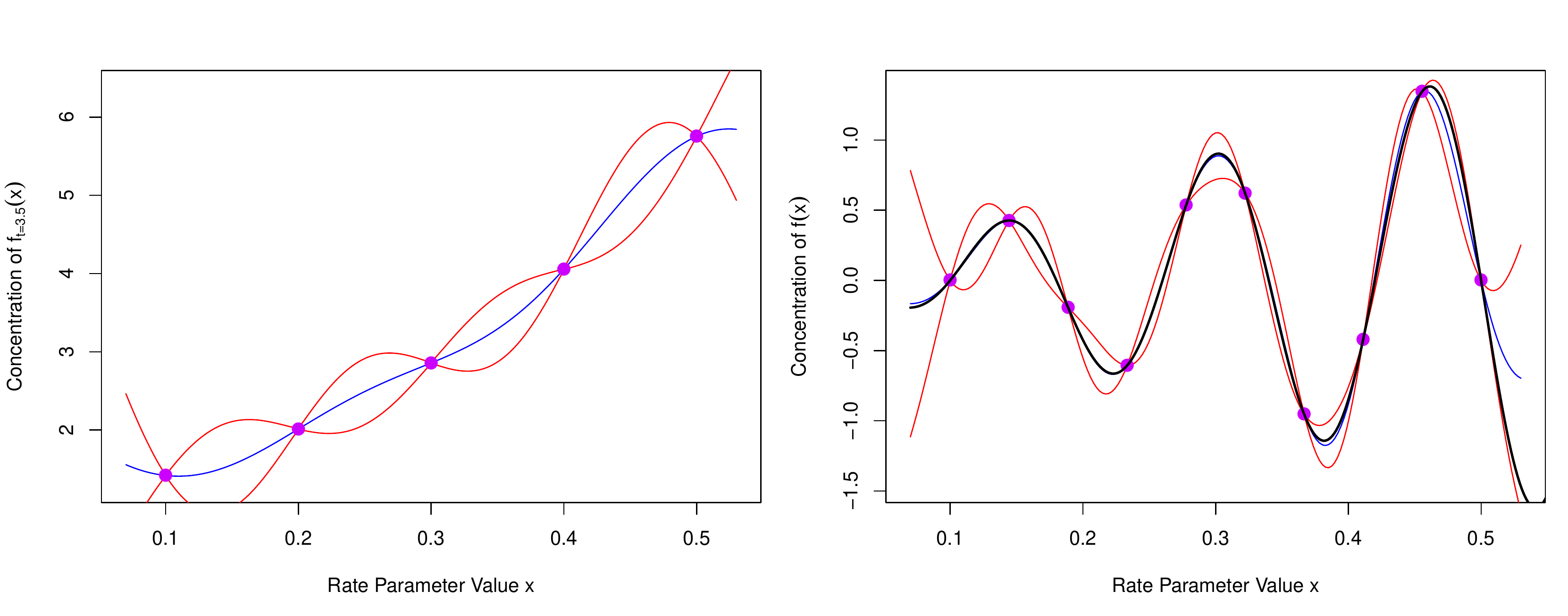} 
\caption{Left panel: an emulator for the simple 1D exponential model. The purple points show five evaluations $D$ of the model at rate 
parameter locations $x^{(j)}$, which are the same as the 5 coloured points in figure~\ref{fig_1d_setup_mod}. The blue line represents the 
emulator's updated expectation ${\rm E}_{D}(f(x))$, and the pair of red lines give the credible interval ${\rm E}_{D}(f(x)) \pm 3 \sqrt{{\rm Var}_{D}(f(x))}$, both as functions of $x$. This defines a region 
between the red lines that we believe is highly likely to contain the true function $f(x)$. Note that evaluation of the emulator is extremely fast, as it only requires matrix multiplication. Right panel: an emulator for the more complex function given by equation~(\ref{eq_comp_emul}). The true function
$f(x)$ is shown as the black line, which lies within the emulator credible intervals at all points.}\label{fig_1d_emulation}
\end{center}
\end{figure*}

We now demonstrate the construction of an emulator for the simple one dimensional exponential model. As there is only one output dimension, $f(x)$ is now a scalar, so we drop the $i$ index from equations~(\ref{eq_emulator}-\ref{eq_BLV}).

Figure~\ref{fig_1d_emulation} (left panel) shows output from such an emulator of the simple model defined by equation~\ref{eq_fxt}. We suppose that only $n=5$ runs of the model have been performed at the locations $x^{(j)}=0.1,0.2,0.3,0.4,0.5$, which are shown as the purple points (these are the same as the 
five coloured points in figure~\ref{fig_1d_setup_mod}). 
We therefore have the model output values 
\begin{align}\label{eq_Di_sim}
D &= (f(x^{(1)}), f(x^{(2)}),\dots,f(x^{(5)}))^T  \\
&= (e^{0.1\times3.5}, e^{0.2\times3.5},\dots,e^{0.5\times3.5})^T \nonumber 
\end{align} 
where again the output of interest has $t=3.5$. 

We use a simplified form of the emulator given by equation~(\ref{eq_emulator}), where we choose the polynomial terms $\beta_{j} g_{j}(x_A)$ to represent only a constant term $\beta_0$. As we only have one input variable, there is no distinction between inactive and active variables so we also set $w(x)$ to zero, and hence the emulator equation~(\ref{eq_emulator}) reduces to 
\be\label{eq_sim_em}
f(x) \;\;=\;\; \beta_0 + u(x)
\ee
For simplicity we treat the constant term $\beta_0$ as known and hence set ${\rm Var}(\beta_0)=0$, and choose prior expectation ${\rm E}(\beta_0)=\beta_0 = 3.5$, a value which we expect the function outputs to be approximately centred around. We specify the parameters in the covariance function for $u(x)$ given by equation~(\ref{eq_corr}) to be $\sigma_{u}=1.5$ and $\theta = 0.14$ representing curves of moderate smoothness: this process will be discussed in more detail for the full Arabidopsis model. 

All expectation, variance and covariance terms on the right hand side of equations~(\ref{eq_BLE}) and (\ref{eq_BLV}) can now be found using equations~(\ref{eq_sim_em}), (\ref{eq_corr}) and (\ref{eq_Di_sim}), for example, 
\ba
{\rm E}(f(x)) &=& \beta_0 \\
{\rm Var}(f(x)) &=& \sigma_{u}^2 \\
{\rm E}(D) &=& (\beta_0, \dots, \beta_0)^T
\ea
while ${\rm Cov}( f(x), D)$ is a row vector of length $n$ with $j$th component
\ba
{\rm Cov}( f(x), D)_j &=& {\rm Cov}( f(x), f(x^{(j)})) \\
&=& \sigma^2_{u} {\rm exp}\left\{- \frac{\|x-x^{(j)}\|^2}{  \theta^2}\right\} \nonumber
\ea
and similarly ${\rm Var}(D)$ is an $n\times n$ matrix with $(j,k)$ element
\ba
{\rm Var}(D)_{jk} &=& {\rm Cov}( f(x^{(j)} ), f(x^{(k)})) \\
&=& \sigma^2_{u} {\rm exp}\left\{- \frac{\|x^{(j)}-x^{(k)}\|^2}{  \theta^2}\right\} \nonumber
\ea
We can now calculate the adjusted expectation and variance ${\rm E}_{D}(f(x))$ and ${\rm Var}_{D}(f(x))$ from equations~(\ref{eq_BLE}) and (\ref{eq_BLV}) respectively.

Figure~\ref{fig_1d_emulation} (left panel) shows ${\rm E}_{D}(f(x))$ as a function of $x$ as the blue line. We can see that it precisely interpolates the five known runs at outputs $D$, which is desirable as $f(x)$ is a deterministic function. The blue line also gives a satisfactory estimate of the true function $f(x) = \exp(3.5x)$. 
The red pair of lines give the credible interval ${\rm E}_{D}(f(x)) \pm 3 \sqrt{{\rm Var}_{D}(f(x))}$ as a function of $x$. This defines a region 
between the lines that we believe is highly likely to contain the true function $f(x)$. Another desirable feature of the emulator is that these credible 
intervals decrease to zero width at the five known run locations, as is appropriate for a deterministic function, as we precisely know the value of $f(x)$ there (and
because we have no inactive variables). Therefore when $x$ is close to a known run we are more certain about the possible values of $f(x)$, compared to when $x$ is far from any such runs.

Figure~\ref{fig_1d_emulation} (right panel) shows an emulator as applied to a more complex 1-dimensional function. Here the true function 
is
\begin{equation}\label{eq_comp_emul}
f(x) \;\;=\;\; 3 \,x \sin\left(\frac{5\pi(x-0.1)}{ 0.4}\right)
\end{equation}
which has been simulated at only 10 input points evenly spread between $x^{(1)} = 0.1$ and $x^{(10)} = 0.5$. Here the prior emulator specifications were as in the previous example, but with ${\rm E}(\beta_0)=0$,  $\sigma_{u}=0.6$ and $\theta = 0.06$ allowing for functions with more curvature, centred around zero. 
As before the blue and red lines show ${\rm E}_{D}(f(x))$ and ${\rm E}_{D}(f(x)) \pm 3 \sqrt{{\rm Var}_{D}(f(x))}$ as functions of $x$. The true function $f(x)$ is given by the solid black line and it can be seen that it lies within the credible region for all $x$, only getting close to the boundary for $x>0.5$. This demonstrates the power of the emulation process: with the use of only 10 points the emulator accurately mimics a reasonably complex function with five turning points. 
We will demonstrate the effectiveness of emulators in higher dimensions for the main Arabidopsis model example.

\subsection*{History matching: an efficient global parameter search}

Bayesian emulation is very useful in a variety of situations. As emulators are extremely fast to evaluate, they can replace the original model in any larger calculation, for example when designing future experiments \cite{Ver_design_sysbio_stats,Ver_design_sysbio_bio}. They can also provide much structural insight into the behaviour of the model. 
One of the most important applications of emulation is to the problem of performing a global parameter search. In this section we describe a powerful iterative global search method known as history matching, which has been \red{successfully employed in a variety of scientific disciplines 
including galaxy formation~\cite{Vernon10_CS,Vernon10_CS_rej,vernon_astro, galf_stat_sci,Vernon:2016aa}, epidemiology~\cite{Yiannis_HIV_1,Yiannis_HIV_2,Yiannis_HIV_3,McCreesh2017}, oil reservoir
modelling~\cite{Craig96_Pressure,Craig97_Pressure,JAC_Handbook,JAC_sma_samp},
climate modelling~\cite{Williamson:2013aa}, environmental science~\cite{asses_mod,Goldstein2016} \blue{and traffic modelling~\cite{Boukouvalas:2014aa}}. Many of these applications involved models with substantial runtime, for which the process of emulation is vital. 
}

When confronting a systems biology model with observed data the following questions are typically asked:
\vspace{-0.2cm}
\bn
\item Are there any input parameter settings that lead to acceptable matches between the model output and observed data?
\item If so, what is the full set $\mathcal{X}$ that contains all such input parameter settings?
\en
\smallskip
History matching is designed to answer these questions. It proceeds iteratively and employs {\it implausibility measures} to determine parts of the input space that can be discarded from further investigation. 

We can ask the question: for an unexplored input parameter setting $x$, how far would the emulator's expected value for the individual function output $f_i(x)$ be from the corresponding observed value $z_i$ before we could deem it highly unlikely for $f_i(x)$ to give an acceptable match were we to evaluate the function at this value of $x$? The implausibility measure $I_i(x)$ captures this concept, and is given by:
\be\label{eq_imp1}
I_i^2(x) \;\;=\;\; \frac{({\rm E}_{D_i}(f_i(x)) - z_i)^2}{ {\rm Var}_{D_i}(f_i(x)) + {\rm Var}(\epsilon_i) + {\rm Var}(e_i)}
\ee
The numerator of equation~(\ref{eq_imp1}) gives the distance between the emulator expectation ${\rm E}_{D_i}(f_i(x))$ and the observation $z_i$, while the denominator standardises this quantity by all the relevant uncertainties regarding this distance: the emulator variance ${\rm Var}_{D_i}(f_i(x))$, the model discrepancy variance ${\rm Var}(\epsilon_i)$ and the observation error variance ${\rm Var}(e_i)$. This structure is a direct consequence of equations~(\ref{eq_zye}) and (\ref{eq_yfep}).
A large value of $I_i(x)$ for a particular $x$ implies that we would be unlikely to obtain an acceptable match between $f_i(x)$ and $z_i$ were we to run the model there. Hence we can discard the input $x$ from the parameter search if $I_i(x)>c$, for some cutoff $c$, \red{and refer to such an input as implausible}.
We may choose the cutoff $c$ by appealing to Pukelsheim's 3-sigma rule \cite{threesigma}, \red{which is the powerful result that states that for {\it any} continuous, unimodal distribution, 95\% of its probability must lie within $\pm 3 \sigma$, regardless of asymmetry, skew, or heavy tails, which suggests} that a choice of $c=3$ could be deemed reasonable 
\cite{Vernon10_CS}. 


We can combine the implausibility measures $I_i(x)$ from several outputs in various simple ways, for example we could maximise over all outputs defining
\be\label{eq_maximp}
I_M(x) \;\;=\;\; \max_{i \in Q} I_i(x)
\ee
where $Q$ represents the collection of all outputs, or some important subset of them (often we will only emulate a small subset of outputs in early iterations). A more robust approach would be to consider the second or third maximum implausibility, hence allowing for some inaccuracy of the emulators~\cite{Vernon10_CS}. Also, multivariate implausibility measures are available (see~\cite{Vernon10_CS} for details), but these require a more detailed prior specification, for example this requires covariances between different components of $e$ and $\epsilon$. \red{Note that a low value of the implausibility $I_M(x)$ does not imply that the input point $x$ is `good' or `plausible'
as it still may lead to a poor fit to outputs that have not been included in $Q$ yet. Also, low implausibility at $x$ may occur because of a high emulator variance ${\rm Var}_{D_i}(f_i(x))$ which once resolved following further runs of the model, may then lead to a high implausibility at $x$. Hence we refer to low implausibility inputs $x$ as ``non-implausible", consistent with the literature \cite{Craig97_Pressure,JAC_sma_samp,Vernon10_CS,galf_stat_sci,Yiannis_HIV_1,Yiannis_HIV_2,Vernon:2016aa}.}

We proceed iteratively, discarding regions of the input parameter space in waves, refocussing our search on the remaining `non-implausible' inputs at each wave.
Prior to performing the $k^{th}$ wave, we define the current set of non-implausible input points as $\mathcal{X}_k$ and the set of outputs that we considered for emulation in the previous wave as~$Q_{k-1}$. We proceed according to the following algorithm. 

\vspace{0.3cm}
\begin{enumerate}
\item Design and evaluate a well chosen set of runs over the current non-implausible space $\mathcal{X}_k$. e.g. using a maximin \red{Latin} hypercube with rejection \cite{Vernon10_CS}. 
\item Check to see if there are new, informative outputs that can now be emulated accurately (that were difficult to emulate well in previous waves) and add them to the previous set $Q_{k-1}$, to define $Q_k$. 
\item Use the runs to construct new, more accurate emulators defined only over the region $\mathcal{X}_k$ for each output in $Q_k$.
\item The implausibility measures $I_i(x)$, $i \in Q_k$, are then recalculated over $\mathcal{X}_k$, using the new emulators.
\item Cutoffs are imposed on the Implausibility measures $I_i(x) < c$ and this defines a new, smaller
non-implausible volume $\mathcal{X}_{k+1}$ which should satisfy $\mathcal{X} \subset \mathcal{X}_{k+1} \subset \mathcal{X}_k$.
\item Unless a) the emulator variances for all outputs of interest are now small in comparison to the other sources of
uncertainty due to the model discrepancy and observation errors, or b) the entire input space has been deemed implausible, return to step~1.
\item If 6 a) is true, generate as large a number as possible of acceptable runs from the final non-implausible volume~$\mathcal{X}$, sampled depending on scientific goal.
\en
\vspace{0.3cm}

We then analyse the form of the non-implausible volume $\mathcal{X}$, the behaviour of model evaluations from different locations within it and the corresponding biological implications. \red{If the entire input space has
been deemed implausible in step 6 b), this may imply that the model is inconsistent with the observed data, with respect to the specified uncertainties. This could be because the biological principles that 
underlie the model's structure are incorrect, and hence remodelling is required. Or that we may have underestimated the observation errors, the model discrepancy or even the emulator uncertainty, although emulator diagnostics~\cite{Tony_EmDiag} combined with the choice of fairly conservative cutoffs should make the latter unlikely. Note that concluding that a far larger model discrepancy is needed is essentially stating that the model is highly inaccurate, and may, for example be judged unfit for purpose.}

The history matching approach is powerful for several reasons:
\vspace{0.3cm}
\bi
\item As we progress through the waves and reduce the volume of the region of input space of interest, we expect the function $f(x)$ to become smoother, and hence to be more accurately approximated by the regression part of the emulator $\beta_{ij} g_{ij}(x_{A_i})$, which is often composed of low order polynomials (see equation~\ref{eq_emulator}).
\item At each new wave we have a higher density of points in a smaller volume and hence the Gaussian process term $u_i(x_{A_i})$ in the emulator will be more effective, as it depends mainly on the proximity of $x$ to the nearest runs. 
\item In later waves the previously strongly dominant active inputs from early waves will have their effects curtailed, and hence it will be easier to select additional active inputs, unnoticed before. 
\item There may be several outputs that are difficult to emulate in early waves (perhaps because of their erratic behaviour in uninteresting parts of the input space) but simple to emulate in later waves once we have restricted the input space to a much smaller and more biologically realistic region.
\ei

\subsubsection*{1-dimensional example}

\begin{figure*}
\vspace{-0.5cm}
\begin{center}
\hspace{-0.0cm}\includegraphics[scale=0.49,angle=0]{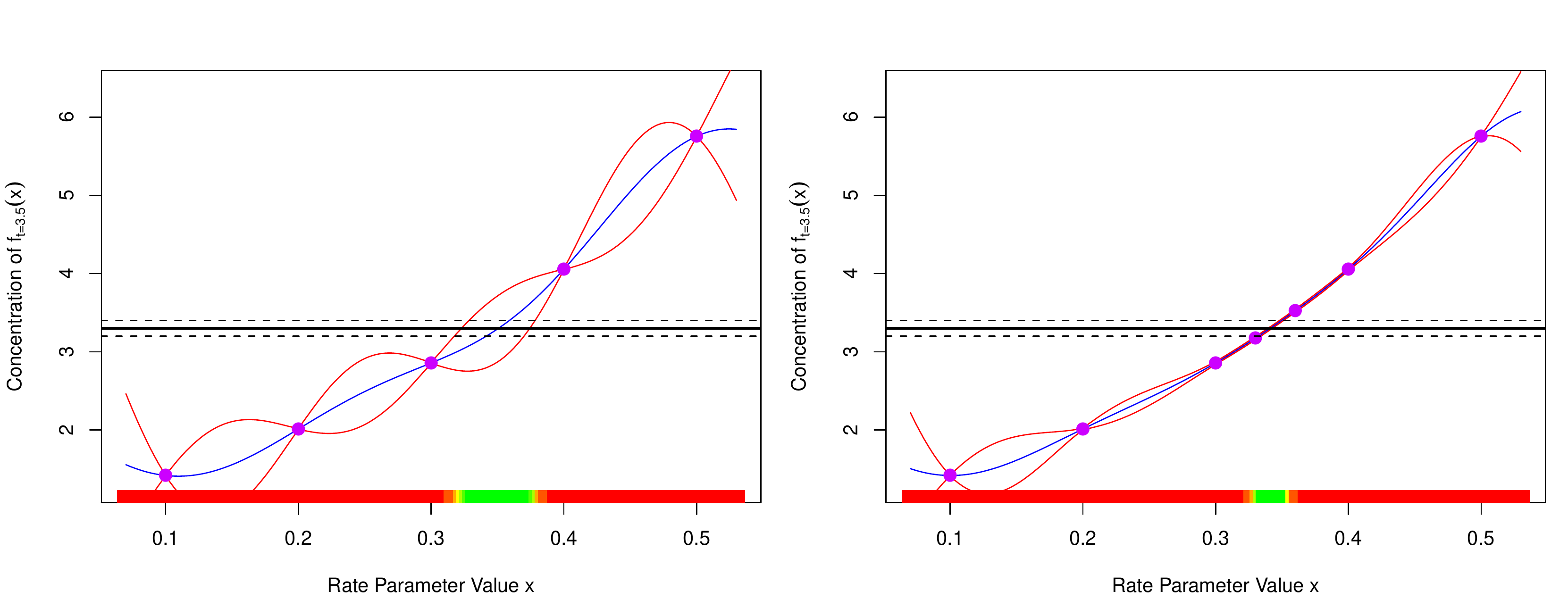} 
\caption{Left panel: the emulator expectation and credible intervals as in figure~\ref{fig_1d_emulation}, however now the observation $z$ plus observed error has been included as the horizontal black solid and dashed lines respectively. 
The implausibilities $I(x)$ are represented by the colours on the x-axis: red, yellow and green for high ($I(x) > 3.5$), borderline ($3.5< I(x) <3$) and low ($I(x) < 3$) implausibility respectively, with the green interval defining the non-implausible region $\mathcal{X}_2$ for the second wave. 
Right panel: the second wave is performed by evaluating two runs located within $\mathcal{X}_2$. The emulator becomes more accurate over 
$\mathcal{X}_2$ and the implausibility more strict, hence defining the smaller non-implausible region $\mathcal{X}_3$, given by the green interval.
As the emulator is now far more accurate that the observed errors within $\mathcal{X}_3$, additional runs will not significantly reduce $\mathcal{X}_3$ any further, and hence the history match is complete.}\label{fig_1d_HM}
\end{center}
\end{figure*}

We now demonstrate the history matching process as applied to the simple 1-dimensional exponential example.
Figure~\ref{fig_1d_HM} (left panel) shows the emulator expectation and credible intervals as in figure~\ref{fig_1d_emulation}, however now the observation $z$ plus observed error has been included as the horizontal black solid and dashed lines respectively. Here we have set the model discrepancy to zero ($\sigma_\epsilon=0$) and reduced the size of the observation errors $\sigma_e$ for clarity. 
Also given are the implausibilities $I(x)$ as represented by the colours on the x=axis: red, yellow and green for high ($I(x) > 3.5$), borderline ($3.5< I(x) <3$) and low ($I(x) < 3$) implausibility respectively.

The non-implausible space  $\mathcal{X}_1$ at wave 1 is the full initial range of the rate parameter $x$, which is $0.075<x< 5.25$. If we impose cutoffs of $I(x)<3$ then this defines the wave 2 non-implausible space $\mathcal{X}_2$ as shown by the green region of the x-axis in figure~\ref{fig_1d_HM} (left panel). 

We then perform the second wave by designing a set of two more runs over $\mathcal{X}_2$, reconstructing the emulator over this region, and recalculating the implausibility measure $I(x)$. The results of this second wave are shown in figure~\ref{fig_1d_HM} (right panel). It can be seen that the emulator is now highly accurate over the 
$\mathcal{X}_2$ region and that the non-implausible region in green has been further reduced. As the emulator is now far more accurate than the corresponding observation error, we may stop the analysis with this wave as $\mathcal{X}_3 \simeq \mathcal{X}$, implying that further runs will do little to reduce the non-implausible region further. Note that providing we have enough runs in each wave, we would often create new emulators at each wave, defined only over the current green non-implausible region \cite{Vernon10_CS}, instead of updating emulators from previous waves, as in figure~\ref{fig_1d_HM}. \blue{See the Supplementary Material for R code to reproduce the example model output, discrepancy, emulation and history matching plots of figures 1,2,4 and 5 respectively.}

\subsection*{\red{History Matching and Bayesian MCMC}}

\red{Here we discuss the standard form of a full Bayesian analysis, and compare it to the above history matching approach, highlighting the relative strengths and weaknesses of each method.}

\red{History matching attempts to answer efficiently some of the most important questions that a modeller may have, identifying if the model can 
match the data, and where in input space such acceptable matches can be found. It requires only a limited specification related to the key 
uncertain quantities, in terms of means, variances and covariances.
A fully Bayesian approach goes further \cite{bernardo2006bayesian}, and delivers a posterior distribution across all uncertain quantities, which  has the benefit of providing probabilistic answers to most scientific questions e.g. in this context it gives the posterior distribution of the location of the true input $x^*$. However, it requires a more detailed  
prior specification of joint probability distributions across all these quantities, and critically, it also assumes the existence of a single true $x^*$ (and the accuracy of the statistical model that defines it). This may not be judged appropriate, for example, in a situation where two different $x^*$s were found that gave good matches to two different subsets of the observed data.
In more detail, a fully Bayesian specification in the context of 
equations~(\ref{eq_zye}) and (\ref{eq_yfep}) requires the multivariate distributions $\pi(z|y), \pi(y|f(x^*)),\pi(f(x^{(j)})|x^{(j)})$ for a collection of inputs $x^{(j)}$, and 
a prior distribution $\pi(x^*)$ over the true input $x^*$. Meaningful specifications of this form can be difficult to make, for example, often familiar 
distributional forms are assumed such as the multivariate normal distribution, but such choices are often made for mathematical convenience or computational tractability. Really, choices of this kind demand a careful justification, without which results such as the posterior for $x^*$ rapidly lose meaning. For example, the fully Bayesian approach will, after substantial calculation, return a posterior for $x^*$, which may be quite narrow, even if the model cannot fit the observed data at all. History matching however may quickly discover this mismatch after a few waves, making further analysis unnecessary.}

\red{The second drawback of the fully Bayesian approach is that it is often hard to perform the necessary calculations, and therefore various 
numerical schemes are required, the most popular being MCMC~\cite{brooks2011handbook}. While MCMC has enjoyed much success, issues 
still remain over the convergence of an MCMC algorithm for even modest dimensional problems~\cite{geyer2011introduction}. Often the likelihood may be highly multimodal, and therefore vast 
numbers of model evaluations are usually required to reach convergence, making MCMC prohibitively expensive for models of even moderate evaluation time. 
In contrast, 
the calculations for history matching are relatively fast and simple.}

\red{A third issue is that of robustness: small changes in the full Bayesian specification, especially involving the likelihood, can lead to substantial 
changes in the posterior. These sensitivities can go unnoticed and can be hard to analyse~\citep{berger1,Berger:2000aa,insua1,BABA_paper1}, 
but will call into question the resulting scientific conclusions. }

\red{Due to these issues, we support the fully Bayesian approach, but only for cases where such detailed calculations are warranted, say for a well 
tested, accurate biological model, which possesses well understood model deficiencies, and which is to be combined with data that have a trusted 
observation error structure, and critically, where full probabilistic results are deemed essential. 
If, instead, the main concern is to check whether the model can fit the data; to see what regions of input parameter space give acceptable matches, 
and for this to be used for further model development, then a history match may be the more appropriate analysis. 
Even if one wishes to forecast the results of future experiments~\cite{Goldstein06_Hatrun,Craig01_Forecasting} and to make 
subsequent decisions~\cite{Ver_design_sysbio_stats}, history matching can be sufficient, as one can either re-weight appropriately the samples generated in the final wave, as 
was done in \cite{Yiannis_HIV_1}, or as stated in step 7 of the history matching algorithm, use more sophisticated sampling in the non-implausible region depending on the scientific question.}

\red{Now the use of emulators can of course facilitate the large number of model evaluations required for Bayesian MCMC algorithms, admittedly at the expense of 
increased uncertainty (see for example \citep{Kennedy01_Calibration,Higdon08a_calibration}). However, a more serious problem is then 
encountered. The likelihood function, which is a core component of Bayesian calculations, is constructed from {\it all} outputs of interest (and 
therefore attempts to describe both the `good' and `bad' inputs simultaneously). Hence we need to be able to emulate with sufficient accuracy {\it all} such outputs, 
including their possibly complex joint behaviour. This may be extremely challenging as often, especially in early waves, there may 
exist several 
erratically behaved outputs that are extremely difficult to emulate, which will dramatically fail emulator diagnostics~\citep{Tony_EmDiag}.
Unfortunately, the likelihood and hence the posterior may be highly sensitive to such poorly constructed emulators.  
Therefore, from a purely practical perspective, employing the full Bayesian paradigm using inadequate emulators constructed at wave 1, 
may be unwise \cite{Vernon10_CS,Vernon10_CS_rej}.}

\subsection*{\red{History Matching as a precursor to MCMC}}

\red{If one does wish to perform a fully Bayesian analysis on a well tested biological model, we would usually recommend performing a history match first~\cite{Vernon10_CS_rej}. This would identify the non-implausible region $\mathcal{X}$ which should contain the vast majority of the posterior. Then MCMC or an equivalent Bayesian calculation (such as importance sampling) can be performed within $\mathcal{X}$, using the accurate emulators defined in the final wave (for an example of this see \cite{Yiannis_HIV_1}).}

\red{This is because if the model is of modest to high dimension, the posterior may often only occupy a tiny fraction of the original input space 
$\mathcal{X}_1$. Unless the model is very fast to evaluate, we would need to use emulators to overcome the MCMC convergence issues, but performing enough model runs to construct sufficiently accurate emulators over the whole of $\mathcal{X}_1$ would be extremely inefficient. 
Iterative history matching naturally provides the accurate emulators that we would need, defined only over $\mathcal{X}$, which should contain the posterior. }

\red{Note that the history match cuts out space efficiently, based on small numbers of easy to emulate outputs in early waves, and designs appropriate runs for the next wave that have good coverage properties for the current non-implausible region. Alternative iterative strategies, 
such as using MCMC at each wave to generate samples from the current posterior (which includes high emulator uncertainty), for use as model runs in the next wave, may be highly 
inefficient and could run into a series of difficulties. Such strategies would not fully exploit the smoothness of the model output, and may tend to cluster points together around the current posterior mode (hence squandering runs, for smooth deterministic models), leading to poor coverage of the full non-implausible space. Such strategies may also be highly misled by inaccurate emulators and the subsequent posterior sensitivity combined with multimodal likelihood issues, leading to clustered designs in the wrong parts of the input space, and in some cases a lack of convergence.}

\red{To conclude, if one really desires a fully Bayesian analysis, then performing a history match first can greatly improve efficiency. In this way we view history matching not as a straight competitor to alternative approaches, but instead a complimentary technique, which has many benefits in its own right. See \cite{Vernon10_CS} and the extended discussion in \cite{Vernon10_CS_rej} for more details of this argument.} 

\subsection*{\blue{History Matching and ABC}}
\blue{Another Bayesian technique that has been developed more recently, and compared (somewhat cautiously) to history matching, by for example~\cite{Holden:aa}, is that of Approximate Bayesian Computation or ABC~\cite{wilkinson:2013}. While the two approaches seem to share some superficial similarities, they are fundamentally different in their goal and the principled way each approach is set up and implemented. For example, ABC attempts to approximate full Bayesian inference and hence to obtain an approximate Bayesian posterior distribution (critically, using a tolerance that tends to zero). History matching is not an inference procedure, as it is simply attempting to rule out all the input space that is clearly inconsistent with the data given the model discrepancy and observation error (which are meaningful `tolerances' that critically will never tend to zero). It is worth noting that if one attempts to specify a meaningful minimum size for the tolerance in ABC, one is arguably not really employing ABC anymore, but is instead just back using Bayesian inference (as shown by~\cite{wilkinson:2013}) in the form of a sampling-resampling algorithm (as described for example by \cite{Smith:1992aa}).
History matching does not attempt to probabilize the remaining input space in any way, which can result in increased efficiency of the parameter search. 
We have directly compared and contrasted the two approaches in~\cite{McKinley:2017aa}, where we demonstrated that a powerful version of ABC failed to find any part of parameter space that matched the observed data for a 22-dimensional stochastic epidemiology model, while history matching found the correct part of input space and many good matches using approximately half the number of runs that the (failed) ABC-SMC approach used.}

\section*{Results}

\subsection*{Application to the hormonal crosstalk network in Arabidopsis root development model}

We now describe the relevant features of the hormonal crosstalk in Arabidopsis root development model~\cite{Liu10_crosstalk}, in preparation for the application of 
the Bayesian emulation and history matching processes introduced above.

The hormonal crosstalk in Arabidopsis model was constructed on the basis of known molecular interactions and experimental evidence, and models the crosstalk between auxin, ethylene and cytokinin via the {\it PLS} gene in Arabidopsis root development. The network for the model is shown in figure~\ref{fig_network} which displays the two main modules of auxin and ethylene signalling. 
\begin{figure}
\vspace{0.cm}
\begin{center}
\hspace{0.0cm} \includegraphics[scale=0.44,angle=0,viewport= 132 35 630 535,clip]{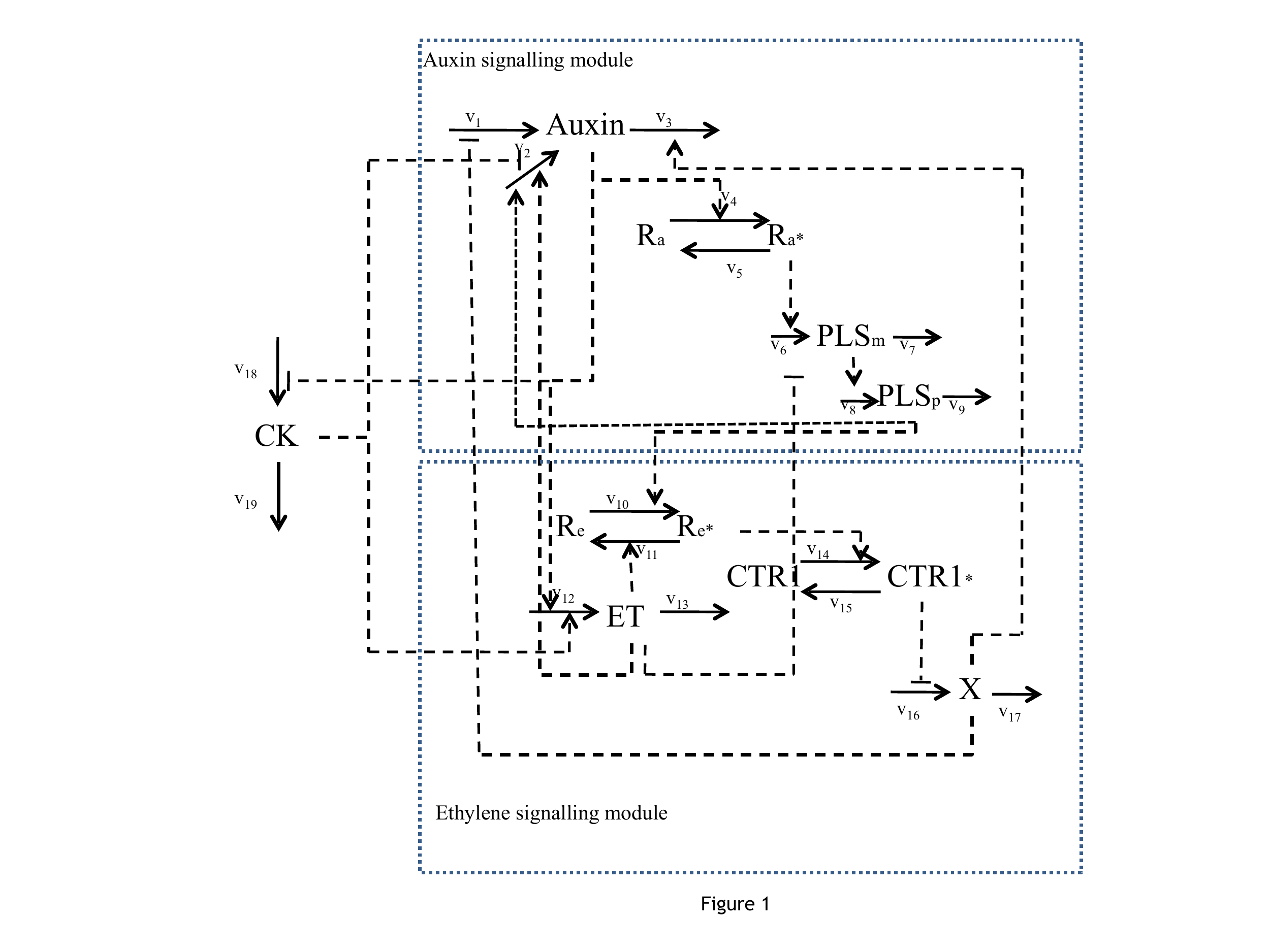}
\caption{
The network for the hormonal crosstalk in Arabidopsis model, 
displaying the two main modules of auxin and ethylene signalling, as described in detail in~\cite{Liu10_crosstalk}.
}\label{fig_network}
\end{center}
\end{figure}
A full description of the model, along with justifications of the various modelling choices employed, can be found in~\cite{Liu10_crosstalk}.

The mathematical representation of the Arabidopsis model, given in table~\ref{tab_difeq}, is a set of 15 ordinary differential equations that 
describe the evolution in time of 15 different biological quantities. Note the analogy with equation~(\ref{eq_fxt}) describing the simple exponential model.
The Arabidopsis model requires the specification of 32 input or rate parameters before it can be evaluated: these are represented in table~\ref{tab_difeq} as the parameters
$(k_1, k_{1a}, \dots, k_{1veth})$. The rate parameter $k_6$ is an exception: it is a control parameter and is set to 0.3 to represent the wildtype 
and 0 to represent the {\it pls} mutant~\cite{Liu10_crosstalk}, and hence it will not be included in our parameter search, leaving 31 free parameters.

As we will compare the model output to data at equilibrium only~\cite{Liu10_crosstalk}, we can perform a substantial dimensional reduction of the input space. Referring to the model equations given in table~\ref{tab_difeq}, we see that 
at equilibrium the derivative on the left hand side of each equation will equal zero, and that the right hand side can hence be rearranged in terms of one less rate parameter. 
For example, the equation for $d[Ra]/dt$ becomes:
\be
 0 = - [Auxin]  [Ra] + \left(\frac{k_5}{k_4} \right) [Ra^*] 
\ee
which depends only on the ratio of $(k_5/k_4)$. Hence data at equilibrium can inform only about the ratio $(k_5/k_4)$
and cannot provide any constraint upon the original parameters $k_4$ and $k_5$ individually (it is worth noting that if the 
model was a stochastic model instead of a deterministic model, it may be possible to learn about the parameters individually, even at equilibrium, 
as discussed in chapter 1 of~\cite{wilkbook}). 
We can therefore remove a total of 8 parameters and reduce the dimension of the input space from 31 to 23, by choosing to work with appropriate rate parameter ratios. The specific rate parameter ratios we use as well as the unaltered rate parameters are given in table~\ref{tab_input_ranges}. Also shown are the 
ranges used to define the initial search space $\mathcal{X}_1$, discussed further below.
\begin{table}
\begin{footnotesize}
\begin{eqnarray}
\vspace{-0.9cm}
\frac{d[Auxin]}{dt}  	&= &  \frac{k_{1a} }{1 + \frac{[X]}{k_1}} + k_2 + \frac{k_{2a} [ET] }{ 1 + \frac{[CK]}{k_{2b}}}  \frac{[PLSp]} { k_{2c} + [PLSp]} - \nonumber \\ \nonumber
\vspace{-0.5cm}    				& &			(k_3 + k_{3a} [X])  [Auxin]         + k_{1vauxin}  [IAA]   \\ \nonumber
	\frac{d[X]}{dt}	&=&  k_{16} - k_{16a}  [CTR1^*] - k_{17}  [X] \\ \nonumber
\frac{d[PLSp]}{dt}   	&=& k_8  [PLSm] - k_9  [PLSp] \\ \nonumber
\frac{d[Ra]}{dt}			&=& - k_4 [Auxin]  [Ra] + k_5  [Ra^*] \\ \nonumber
\frac{d[Ra^*]}{dt} 	&=& k_4  [Auxin] [Ra] - k_5  [Ra^*] \\ \nonumber
\frac{d[CK]}{dt}			&=& \frac{k_{18a} }{ 1 + \frac{[Auxin]}{k_{18}}} - k_{19}  [CK]     + k_{1vCK}  [cytokinin] \\ \nonumber
\frac{d[ET]}{dt}			&=& k_{12} + k_{12a}  [Auxin]  [CK] - k_{13}  [ET]      + k_{1veth}  [ACC] \\ \nonumber
\frac{d[PLSm]}{dt}		&=& \frac{k_6  [Ra^*]} { 1 + \frac{[ET]}{k_{6a}}} - k_7  [PLSm] \\ \nonumber
\frac{d[Re]	}{dt}		&=& k_{11}  [Re^*]  [ET] - (k_{10} + k_{10a}  [PLSp])  [Re] \\ \nonumber
\frac{d[Re^*]}{dt} 	&=& -k_{11}  [Re^*]  [ET] + (k_{10} + k_{10a}  [PLSp])  [Re] \\ \nonumber
\frac{d[CTR1]}{dt}		&=& -k_{14}  [Re^*]  [CTR1] + k_{15}  [CTR1^*] \\ \nonumber
\frac{d[CTR1^*]}{dt} 	&=& k_{14}  [Re^*]  [CTR1] - k_{15}  [CTR1^*] \\ \nonumber
\frac{d[IAA]	}{dt}	&=& 0, \quad \quad \frac{d[cytokinin]}{dt}	\;\;=\;\; 0, \quad \quad \frac{d[ACC]}{dt}  \;\;=\;\; 0  \nonumber
\end{eqnarray}
\end{footnotesize}
\caption{The hormonal crosstalk in Arabidopsis root development model differential equations. See~\cite{Liu10_crosstalk} for details. \blue{See also table 15 in the
Supplementary Material for the dimensions or units of each of the rate constants.}}\label{tab_difeq}
\end{table}
\begin{center}
\begin{table}
\begin{tabular}{|c||c|c|}
\hline
Input Rate & ~~~~~Minimum~~~~ & Maximum \\
Parameters & &    \\
\hline
$k_1  $ &    0.1 &10 \\
$k_2/k_{1a}   $ &   0.02 & 2 \\
$k_{2a}/k_{1a}   $ &  0.28 &28\\
$k_{2b} $  &   0.1 &10 \\
$k_{2c}$    &  1$\times$10$^{-6}$ & 1\\
$k_3/k_{1a}   $   & 0.2 &20 \\
$k_{3a}/k_{1a} $   &  0.045  &4.5\\
$k_5/k_4   $  &  0.1 &10\\
$k_6  $ & \multicolumn{2}{c|}{Control: 0 ({\it pls} mutant) or 0.3 (wildtype)}  \\
$k_{6a} $ &    0.002 & 2000\\
$k_7  $ & 0.1 &10 \\ 
$k_9/k_8  $ & 0.1 &10 \\
$k_{10a}/k_{10} $&    166  &1.66$\times$10$^{4}$ \\
 $k_{11}/k_{10}   $&   166 &1.66$\times$10$^{5}$\\
$k_{12a}/k_{12} $ &   0.1 & 10\\
$k_{13}/k_{12}  $ &   1& 1000 \\
$k_{15}/k_{14}   $ &  2.83$\times$10$^{-4}$  &0.283 \\
$k_{16a}/k_{16}   $&  0.33 & 33.3 \\
$k_{17}/k_{16}    $ &0.033  &3.33\\
$k_{18}     $ &0.01 & 10 \\
$k_{19}/k_{18a}     $ &0.01 &10 \\
$k_{1vauxin}/k_{1a}$ &0.1 &100\\
$k_{1vCK}/k_{18a}$   & 0.1 &10 \\
$k_{1veth}/k_{12} $ & 1 &100 \\
\hline
\end{tabular}
\vspace{0.2cm}
\caption{The input or rate parameter ranges that define the initial search region $\mathcal{X}_1$ over which the history match is performed. Due to symmetries in the model at equilibrium, only ratios of certain parameters will be constrained, hence we choose to work directly with
these ratios, as given in the left column. Note that $k_6$ is a control parameter used to define wildtype or {\it pls} mutant, and hence is not included in the parameter search.}\label{tab_input_ranges}
\end{table}
\end{center}

As we consider ranges of the rate parameters and their ratios which are always positive and span two or more orders of magnitude, we choose to convert to a log scale. Hence we define the 23-dimensional vector $x$ of input parameters for the model as:
\be
x=(\log(k_1),\log ( k_2/k_{1a} ),\dots, \log ( k_{1veth}/k_{12} )) \label{eq_loginputs}
\ee
which corresponds to the first column of table~\ref{tab_input_ranges}, without the inclusion of the control parameter $k_6$. It is this vector of inputs $x$
that will be used in the emulator equations~(\ref{eq_emulator}), (\ref{eq_corr}), (\ref{eq_BLE}), (\ref{eq_BLV}), and that is directly analogous to the 1-dimensional input $x$ 
of the simple model given in equations~(\ref{eq_fxt}) and (\ref{eq_fxt_an}). 
The Arabidopsis model also requires initial conditions for each of the 15 model outputs~\cite{Liu10_crosstalk}, and the values used are given in table~\ref{tab_initial_cond}.

\begin{center}
\begin{table}
\begin{tabular}{|l|c|c|}
\hline
Model   & Initial & Measurement \\
Output &  Concentration & Available \\
\hline
Auxin & 0.1  & Yes\\
X & 0.1&  \\
PLSp & 0.1 &  \\
Ra & 0 & \\
Ra* & 1 & \\
CK & 0.1 &  Yes \\
ET & 0.1 &  Yes \\
PLSm & 0.1& Yes \\
Re & 0& \\
Re* & 0.3& \\
CTR1 & 0& \\
CTR1* & 0.3& \\
IAA & 0 or 1 & \\				
cytokinin & 0 or 1& \\
ACC & 0 or 1& \\
\hline
\end{tabular}
\vspace{0.2cm}
\caption{The list of 15 original model outputs, their initial conditions and whether measurements are available. For simplicity of terminology, exogenous application of IAA, cytokinin or ACC is referred to as "feeding auxin, cytokinin or ethylene".The values of 0 or 1 for IAA, cytokinin and ACC correspond to no feeding or feeding of auxin, cytokinin or ethylene respectively. See~\cite{Liu10_crosstalk} for details.}\label{tab_initial_cond}
\end{table}
\end{center}

We are primarily interested in the behaviour of the four measurable outputs: $[Auxin]$, $[PLSm]$, ethylene $[ET]$ and cytokinin (represented as $[CK]$ in the model). These were measured for the following cases: wild type~(wt), {\it pls} mutant~(mu), wild type fed auxin~(fa), wild type fed ethylene~(fe), wild type fed cytokinin~(fc) and {\it pls} mutant fed ethylene~(mu\_fe). 
The critical behaviour that we want the Arabidopsis model to capture is that of the trends exhibited between certain pairs of measurements. 
For example, the auxin level is seen to decrease in the {\it pls} mutant compared to that of the wild type, while it is seen to increase when ethylene is fed to the wild type compared to the wild type with no feeding. 
A summary of the 16 experimental trends that were used in this analysis is given in table~\ref{tab_trends_mod} (see \cite{Liu10_crosstalk} for details). 
The six different experimental scenarios are correspondingly represented in the model by choosing certain values for the control parameter $k_6$ (which corresponds to the effect of the {\it PLS} gene) and the initial conditions for IAA, ACC and cytokinin, which represent the concentration of feeding chemicals present. The wild type and {\it pls} mutant cases correspond to setting $k_6 = 0.3$ and $k_6 = 0$ respectively, while no feeding implies IAA=ACC=cytokinin=0, with IAA=1, ACC=1 or cytokinin=1 corresponding to the feeding of auxin, ethylene or cytokinin respectively (see table~\ref{tab_trends_mod}).
\begin{center}
\begin{table*}
\begin{tabular}{|l|cccc|c|}
\hline
  	& \multicolumn{4}{c|}{Trend relative to wild type with no feeding} &  Trend relative to {\it pls} mutant with no feeding \\
  	& \multicolumn{4}{c|}{ ($k_6 =0.3$, IAA=ACC=cytokinin=0)}  & ($k_6 =0$, IAA=ACC=cytokinin=0) \\
	\hline
Chemical & {\it pls} mutant & Feed Auxin  & Feed Ethylene & Feed Cytokinin & {\it pls} mutant $+$ Feed Ethylene \\ 
 Output 	& ($k_6=0$)  & (IAA=1) & (ACC=1) & (cytokinin=1) &   ($k_6=0$ and ACC=1) \\
\hline
Auxin &  Down & Up & Up & Down & Down \\
PLSm &  - & Up & Down & Down & -\\
ET   &    No change & Up & Up & Up & -\\
CK    &   Up & Down & Down & Up & -\\
\hline
\end{tabular}
\vspace{0.2cm}
\caption{Summaries of the direction of observed trends of the four measurable chemicals, relative to wild type for the four types of experiment: {\it pls} mutant, feeding auxin, ethylene and cytokinin respectively (first four columns). The final column gives the trend for the case of feeding ethylene to the {\it pls} mutant, relative to the {\it pls} mutant with no feeding. See the text and also~\cite{Liu10_crosstalk} for more detail on the size and related uncertainties for each of the measured trends.}\label{tab_trends_mod}
\end{table*}
\end{center}

To represent the possible model outputs corresponding to each of the cases, we define the time dependant function $h$:
\ba
\!\!\! h_{j,a}(x,t), & a \in &  \!\!\!\!\! \{  \text{wt}, \text{mu},  \text{fa},      \text{fe},      \text{fc},     \text{mu\_fe}  \}   \nonumber   \\
 			& j \in & \!\!\!\!\!  \{ Auxin, PLSm,  ET,    CK   \} \nonumber \\
			& x = &  \!\!\!\!  (\log(k_1),\dots, \log ( k_{1veth}/k_{12} )) \nonumber 
\ea  
where we have introduced a control parameter $a$ that represents the combined choice of plant type and feeding action, the subscript $j$ indexes each of the four measurable chemicals, the vector $x$ represents the vector of rate parameters as before and $t$ represents time.

We are mainly interested in matching the observed trends which are often specified as ratios to wild type. Therefore we choose to work with the log ratio of model outputs, as these will be more robust and allow multiplicative error statements. We also equate these trends to the output of the model at equilibrium~\cite{Liu10_crosstalk}, that is for $ t \rightarrow \infty $, 
and hence we define the main outputs of interest to be 
\be
f_i(x) \; = \;   \lim_{t \rightarrow \infty} \log \left\{ \frac{h_{j,a_2}(x,t)}{h_{j,a_1}(x,t)} \right\} \label{eq_mod_outputs}
\ee
where the subscript $i$ indexes the elements of the list $\{j,a_1,a_2\}$ corresponding to the 16 trends that were actually measured, as presented in table~\ref{tab_trends_mod}.
It is this function $f_i(x)$ that will be directly compared to the observed trends. Again, note the analogy with $f_t(x)$ as defined by equations~(\ref{eq_fxt}) and (\ref{eq_fxt_an}).
We also append to $f_i(x)$ two additional outputs of interest which are not ratios: $\log(h_{auxin,wt}(x,t))$ and $\log(h_{CK,wt}(x,t))$, again evaluated as $t \rightarrow \infty$. These will ensure the acceptable matches found will not have unrealistic concentrations of auxin and cytokinin. 
Note that the Bayesian emulation and history matching methods we propose could be applied to outputs at any time point, and not just to the equilibrium points of primary interest here \red{(see for example~\cite{Yiannis_HIV_1,Yiannis_HIV_2,JAC_Handbook,JAC_sma_samp,Vernon:2016aa})}.

The primary question that the modeller may ask at this point is whether the outputs of the model, in the form of $f_i(x)$, match the observed trends given in table~\ref{tab_trends_mod}, to within an acceptable level of tolerance, and what is the set $\mathcal{X}$ of all rate or input parameters corresponding to such acceptable matches. 

The initial input space $\mathcal{X}_1$ that we choose to perform the global parameter search or history match over is defined in table~\ref{tab_input_ranges}. This was constructed by specifying ranges on the 23 inputs that covered at least one order of magnitude above and below the
single input parameter setting found in~\cite{Liu10_crosstalk}. The ranges of some parameters of particular interest were subsequently increased to allow a wider exploration.
This means we will explore a biologically plausible space that covers at least two orders of magnitude in every dimension, centred (on a log scale) around the original parameter point. This gives rise to a large space $\mathcal{X}_1$, of suitable size to demonstrate our methodology. Note that we could 
make these ranges wider still if this was deemed plausible, which would simply result in us having to perform more waves to complete the history match.

\subsection*{Linking the Arabidopsis model to reality}

The next task is to link the Arabidopsis model formally to reality~\cite{Vernon10_CS,Vernon10_CS_rej,galf_stat_sci}. The Bayesian paradigm allows us 
to represent scientific judgements as probabilistic specifications or, if we follow the Bayes Linear approach, as expectation and variance statements~\cite{Goldstein07_BayesLinearBook}. 
As we do not have access to the precise quantitative values for the observations $z_i$ that feature in equation~\ref{eq_zye}, we instead propose 
values for the observations, observation errors ${\rm Var}(e_i)$ and model discrepancy 
${\rm Var}(\epsilon_i)$ that are consistent both with the observed trends given in table~\ref{tab_trends_mod} and with expert judgement concerning the accuracy of the model and the relevant experiments. We do this for two reasons: firstly to demonstrate that our approach can be reasonably applied to situations where only qualitative data is available, and secondly to highlight what kinds of analysis are possible if quantitative measurements are actually available across all the outputs of interest, hence motivating more detailed future data collection. 
There are several possible ways to assess these quantities while conserving consistency with the observed trends. We choose a conservative, minimal approach,
and specify for the ``Up", ``Down" and ``No Change" trends that $z_i = 1.24, -1.24$ and 0, and that $\sigma_i= 0.35, 0.35$ and $0.061$ respectively, where $\sigma_i$ represents the combined model discrepancy and observed errors 
\be\label{eq_sigma_comb_md_obs}
\sigma_i \;\;=\;\; \sqrt{ {\rm Var}(\epsilon_i) + {\rm Var}(e_i)}.
\ee
These combined specifications have been made so that the intervals 
\be\label{eq_intz}
z_i \: \pm \: 3 \: \sigma_i
\ee
represent an increase of between 20\% to ten fold for the ``Up" trends, a decrease also of between 20\% to ten fold for the ``Down" trends, and an 
interval of 40\% decrease to 40\% increase for the ``No Change" trend. These intervals, assumed symmetric on the log scale, were formulated by answering the natural question: where would each 
model output have to lie to avoid violating the trends given in table~\ref{tab_trends_mod}, considering relevant observational and model uncertainties? 
This specification captures the main features of the trend data and is sufficient for our purposes of demonstrating the Bayesian history matching methodology. 
Obviously, a more detailed treatment would involve having more information regarding the observations $z_i$ themselves, and their associated measurement errors represented by ${\rm Var}(e_i)$. Also, were we to consider in more detail 
the known deficiencies of the model, we could give a more detailed specification of the model discrepancy ${\rm Var}(\epsilon_i)$, \blue{which would most likely include correlations between different outputs that exploited the joint structure suggested by the choice of chemical, choice of mutant and choice of feeding regime, or even including a simple 
dependence on certain input parameters}. 
See \cite{Vernon10_CS}, \cite{asses_mod} and \cite{Craig97_Pressure} for examples of more detailed model discrepancy specifications in alternative applications, and \cite{Goldstein09_Reify} and \cite{Vernon10_CS_rej} for further discussions.

Figure~\ref{fig_outputs_1d_w1} shows all 16 intervals corresponding to the measured trends, as represented by equation~(\ref{eq_intz}), given as the black error bars, on a log scale. Also shown (as the first two errors bars from the left) are the two additional non-ratio wildtype outputs for Auxin and Cytokinin, 
which are given reasonably wide intervals of 0.24 plus or minus an order of magnitude~\cite{Liu10_crosstalk}. 

The specification of $z_i, {\rm Var}(e_i)$ and ${\rm Var}(\epsilon_i)$ or equivalently $\sigma_i$, can be used to define an `acceptable match' between model output and observed data via the implausibility measures of equation~(\ref{eq_imp1}),
as any model evaluation that satisfies $I_i(x) < c$ for some cutoff $c$. A common choice is $c=3$, based on 
Pukelsheim's 3-sigma rule (see~\cite{threesigma}). 
We may impose this constraint simultaneously across all of the 18 outputs shown in figure~\ref{fig_outputs_1d_w1}, by demanding that $I_M(x)<c$ where $I_M(x)$ is the maximum implausibility defined by equation~(\ref{eq_maximp}), or we could impose a less stringent criteria by constraining the second or third maximum implausibility instead, which would allow model runs to deviate from one or two outputs respectively.

\begin{figure*}
\vspace{0.cm}
\begin{center}
\hspace{0.cm} \includegraphics[scale=0.78,angle=0]{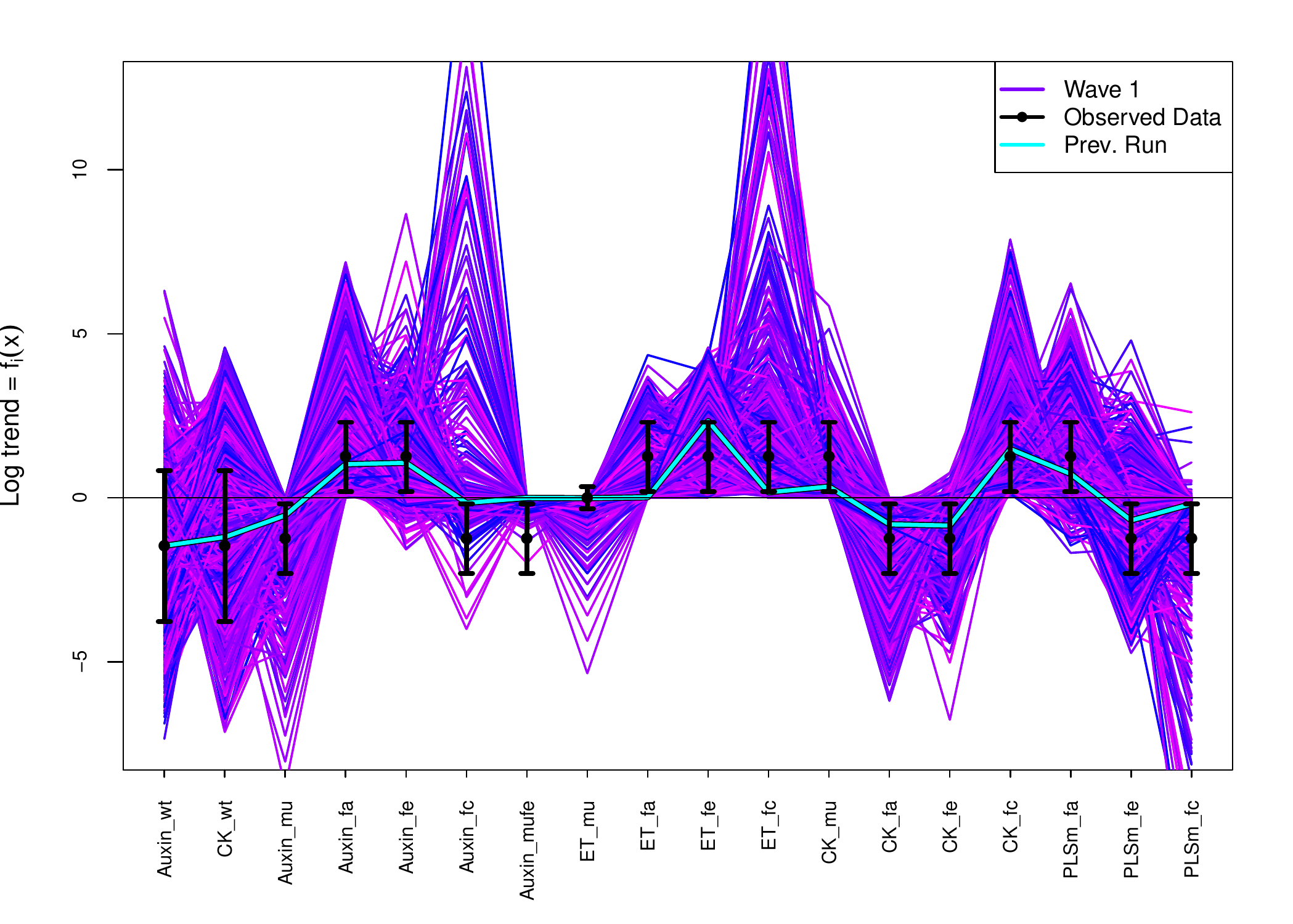}
\caption{The 2000 wave 1 run outputs $f_i(x)$ for all 18 outputs considered (see equation~(\ref{eq_mod_outputs})) are shown as the purple lines, with the observed data $z_i \pm 3 \sigma_i$ given as the black error bars, and the 
best run previously found by~\cite{Liu10_crosstalk} shown as the light blue line. The horizontal black line at zero represents no trend. As these runs were generated from a space filling maximin Latin hypercube design, they can give 
substantial insight into the broad behaviour of the model over the initial search region $\mathcal{X}_1$.
We can see that some outputs are seemingly constrained to give only positive (e.g. Auxin\_fa) or negative (e.g. Auxin\_mu) trends, and that many of the runs are far from the target ranges (as the y-axis is on a log scale).
We also find that no individual wave 1 run passes through every one of the target intervals.}\label{fig_outputs_1d_w1}
\end{center}
\end{figure*}

\subsection*{Bayesian emulation of the Arabidopsis model}
We can now proceed with the first wave of emulation of the Arabidopsis model as follows. \red{Note that several packages are available 
that perform standard Gaussian Process emulation \blue{(see for example the BACCO~\cite{Hankin:2005aa} and GPfit~\cite{JSSv064i12} packages in R, or GPy~\cite{gpy2014} for Python) which may be of use to the uninitiated, as an alternative to the slightly more sophisticated emulators we describe here.}}

First we design a set of 2000 wave 1 runs over the initial search region $\mathcal{X}_1$
based on a maximin \red{Latin} hypercube design (see figure~\ref{fig_lhc} and \cite{Santner03_DACE,SWMW89_DACE}), using for 
example the lhs() function in R~\cite{R-Core-Team:2015aa}. Each of these runs specifies a distinct set of values of all the rate parameters in $x$, and therefore for each run the differential equations given in table~\ref{tab_difeq} were solved numerically using the lsoda() function again in R, with initial conditions given in table~\ref{tab_initial_cond}, up to $t=10000$ seconds to ensure 
equilibrium is reached (equilibrium was then checked). Each run took approximately 1 second of real time to evaluate, implying that although this is a relatively fast model, it is still too slow to exhaustively search the full 23 dimensional input space, which would likely require a vast number of runs. The emulators that we develop turn out to be 4 orders of magnitude faster than the model, and hence allow a much more detailed and efficient exploration. This ratio of emulator speed versus model speed actually improves as the model complexity increases, as the speed of an emulator is a function of the number of runs used to construct it~\cite{Vernon10_CS}. \red{Note that when choosing the number of wave 1 runs, the computer model literature tentatively suggests that at least 10$d$ are required for emulator construction, where $d$ is the dimension of the input space. Of course, depending on the complexity of the model, far more may be needed. Here, as the Arabidopsis model is of reasonable speed, we could afford to run 2000 runs per wave, and this allows the fitting of higher order polynomial terms such as cubics, once a restricted set of active inputs has been identified. Also, 2000 runs allows for a tractable inverse ${\rm Var}(D_i)^{-1}$ that is computed in the emulator equations~(\ref{eq_BLE}) and (\ref{eq_BLV}).}

The wave 1 run outputs $f_i(x)$ for all 18 outputs considered (see equation~(\ref{eq_mod_outputs})) are shown in figure~\ref{fig_outputs_1d_w1} as the purple lines, with the observed data intervals $z_i \pm 3 \sigma_i$ given as the black error bars, and the 
best run previously found by and discussed in~\cite{Liu10_crosstalk} shown as the light blue line. 
As these runs were generated from a space filling design, they can give 
substantial insight into the broad behaviour of the model over the initial search region $\mathcal{X}_1$.
We can see that some outputs are seemingly constrained to give only positive (e.g. Auxin\_fa) or negative (e.g. Auxin\_mu) trends, and that many of the runs are far from the target ranges (as the y-axis is on a $\log_e$ scale).
We also find that no individual wave 1 run passes through every one of the target intervals.
This all suggests that the volume of the non-implausible space $\mathcal{X}$ containing only acceptable runs may be small or indeed zero, and hence we may need several waves for the history match. 

We employ the more general emulator structure as represented by equation~(\ref{eq_emulator}).
For each output $f_i(x)$, we identify the list of active input parameters $x_{A_i}$ by fitting first order polynomials in $x$ and selecting the active inputs based on AIC criteria \blue{(using for example the lm() and step() functions in R~\cite{R-Core-Team:2015aa})}.
We choose the set of deterministic functions $g_{ij}(x_{A_i})$ by selecting terms from the complete third order polynomials in the active inputs, discarding terms again based on AIC criteria (see~\cite{Vernon10_CS, Vernon10_CS_rej, galf_stat_sci, vernon_astro} for more details). 
\red{We show the structure of these wave 1 emulators in terms of the deterministic functions $g_{ij}(x_{A_i})$, and the choice of active variables $x_{A_i}$, in the Supplementary Material}. 
Due to the large number of runs and in the absence of strong prior information, we set ${\rm E}(\beta_{ij}) =0$ and take a large ${\rm Var}(\beta_{ij})$ limit. The $\beta_{ij}$ terms will hence behave, after the Bayes Linear update represented by equations~(\ref{eq_BLE}) and (\ref{eq_BLV}), approximately like their Ordinary Least Squares linear model fits (see~\cite{Vernon10_CS} for details). \blue{Note that the linear models formed at this point, without the inclusion of the Gaussian process part below, would already give reasonably effective emulators (see for example~\cite{Yiannis_HIV_3}).}

We choose the combination of the Gaussian process and nugget variances $\sigma^2_{c_i} = \sigma^2_{u_i} + \sigma^2_{w_i}$ to be equal to the residual standard error from the OLS linear model fit \cite{Vernon10_CS}, and set $\sigma^2_{w_i} = p \sigma^2_{c_i}$ where $p$ is a parameter governing the proportion of variance explained by the inactive variables, taken to be between 0.05 to 0.1, and checked with emulator diagnostics described below. Note that we could design more runs that vary the inactive variables to assess $p$ more accurately, as is done in~\cite{Vernon10_CS}.
We utilise the argument presented in full in~\cite{Vernon10_CS} for choosing correlation lengths for emulators that contain low order polynomials, which states that as we are fitting third order polynomials in the active inputs, we expect the residual surface to look approximately like a fourth (or higher) order surface, and hence we can choose the correlation length accordingly. We hence set the (scaled) correlation lengths $\theta_i$, required for equation~(\ref{eq_corr}), to be 0.35, in agreement with~\cite{Vernon10_CS}, where the inputs $x$ have all been scaled to the range $[-1,1]$. \blue{Note that one can go further and estimate the individual correlation lengths $\theta_i$ using say maximum likelihood, 
which may improve the emulators accuracy, but provided the polynomial surface is fitting well, as judged say by the adjusted $R^2$ of the linear model, this improvement would only be modest.}

Finally, we constructed the emulators by using the Bayes Linear update equations~(\ref{eq_BLE}) and (\ref{eq_BLV}) to obtain ${\rm E}_{D_i}(f_i(x))$ and ${\rm Var}_{D_i}(f_i(x))$ for each output $i$, where $D_i$ is the corresponding vector of 2000 run output values. Emulator diagnostics were then performed by evaluating 200 new diagnostic runs and comparing them to the corresponding emulator predictions, in the form of prediction intervals 
${\rm E}_{D_i}(f_i(x)) \pm 3\sqrt{{\rm Var}_{D_i}(f_i(x))}$ (\red{here 200 runs was deemed sufficient but} see~\cite{Tony_EmDiag} for detailed emulator diagnostics).
In the first wave we found that 13 out of the 18 outputs were straightforward to emulate, in that their emulators were of sufficient accuracy to allow reasonably large parts of the input space to be removed, while simultaneously satisfying emulator diagnostics.
The remaining 5 outputs were left to be considered in later waves. 
Each of these 13 outputs (that define $Q_1$) required between 7-13 active inputs $x_{A_i}$, out of a total of 31 full or 23 reduced input parameters $x$, which represents a substantial dimensional reduction and hence a large benefit to the emulator construction process and subsequent parameter search, as discussed in~\cite{Vernon10_CS_rej}. This is in addition to the speed increase of using emulators as they are in this case $10^4$ times faster to evaluate than the full Arabidopsis model. Note that each one of the 23 inputs featured in at least one of the 13 emulators.  

\subsection*{History matching the Arabidopsis model}

We now employ the iterative history matching strategy described above to the Arabidopsis model. 

As well as the maximised implausibility $I_M(x)$ defined by equation~(\ref{eq_maximp}), we also use the more robust second and third maximum implausibilities denoted $I_{2M}(x)$ and $I_{3M}(x)$ respectively, defined using the set $Q_k$ of outputs considered in wave $k$, as these implausibility measures are more robust to emulator failure. 
In the first wave, only $I_{2M}(x)$ and $I_{3M}(x)$ were used with conservative cutoffs $c$ of 3.25 and 3 imposed respectively. 
This defined $\mathcal{X}_2$: the non-implausible space remaining 
after wave 1, which had a volume of $2.06\times10^{-1}$ of the original input space $\mathcal{X}_1$.
\begin{figure*}
\vspace{0.cm}
\begin{center}
\begin{tabular}{lll}
\hspace{-0.65cm} \includegraphics[scale=0.421,angle=0,viewport= 2 2  405 310,clip]{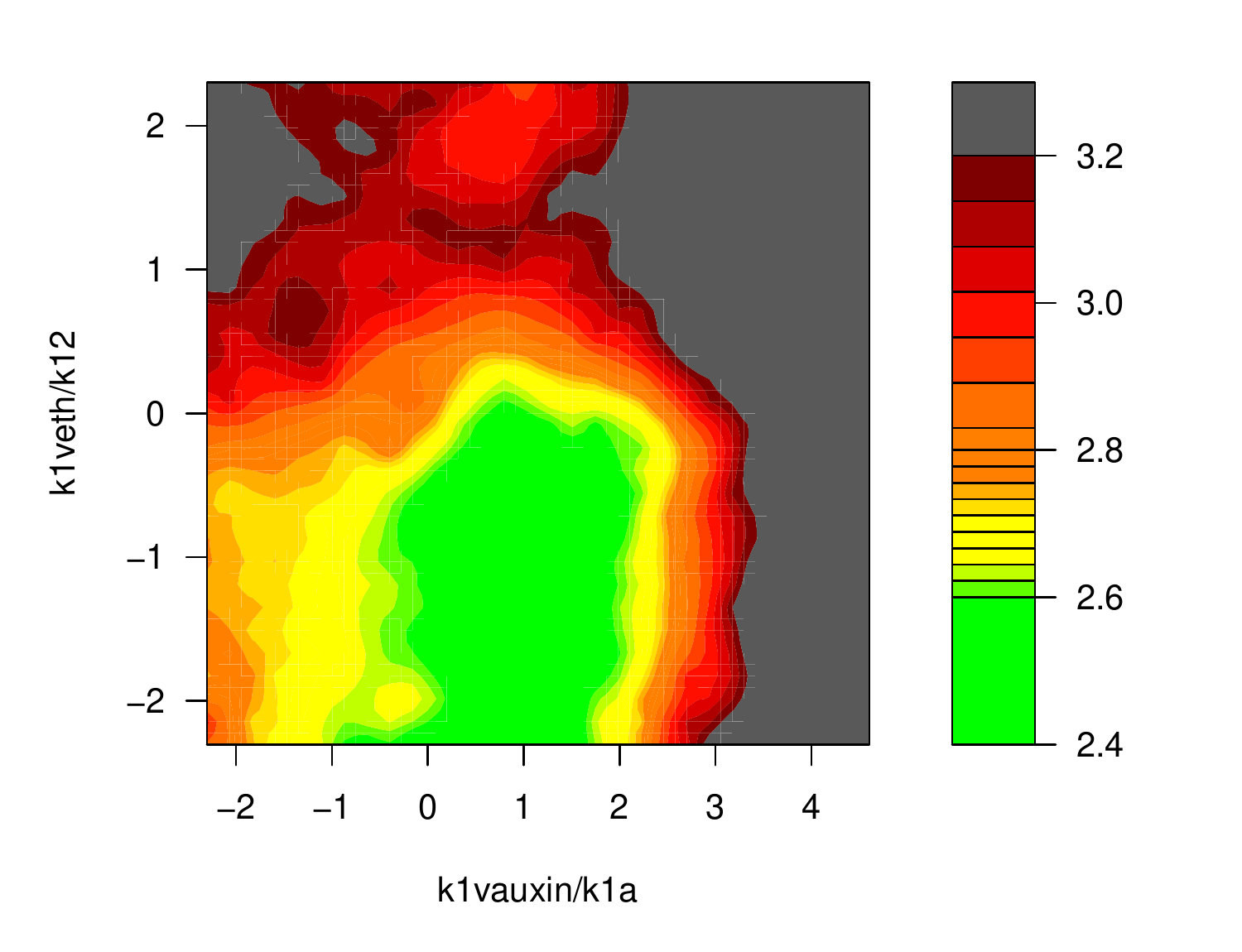} &
\hspace{-0.5cm} \includegraphics[scale=0.421,angle=0,viewport= 42 2 405 310,clip]{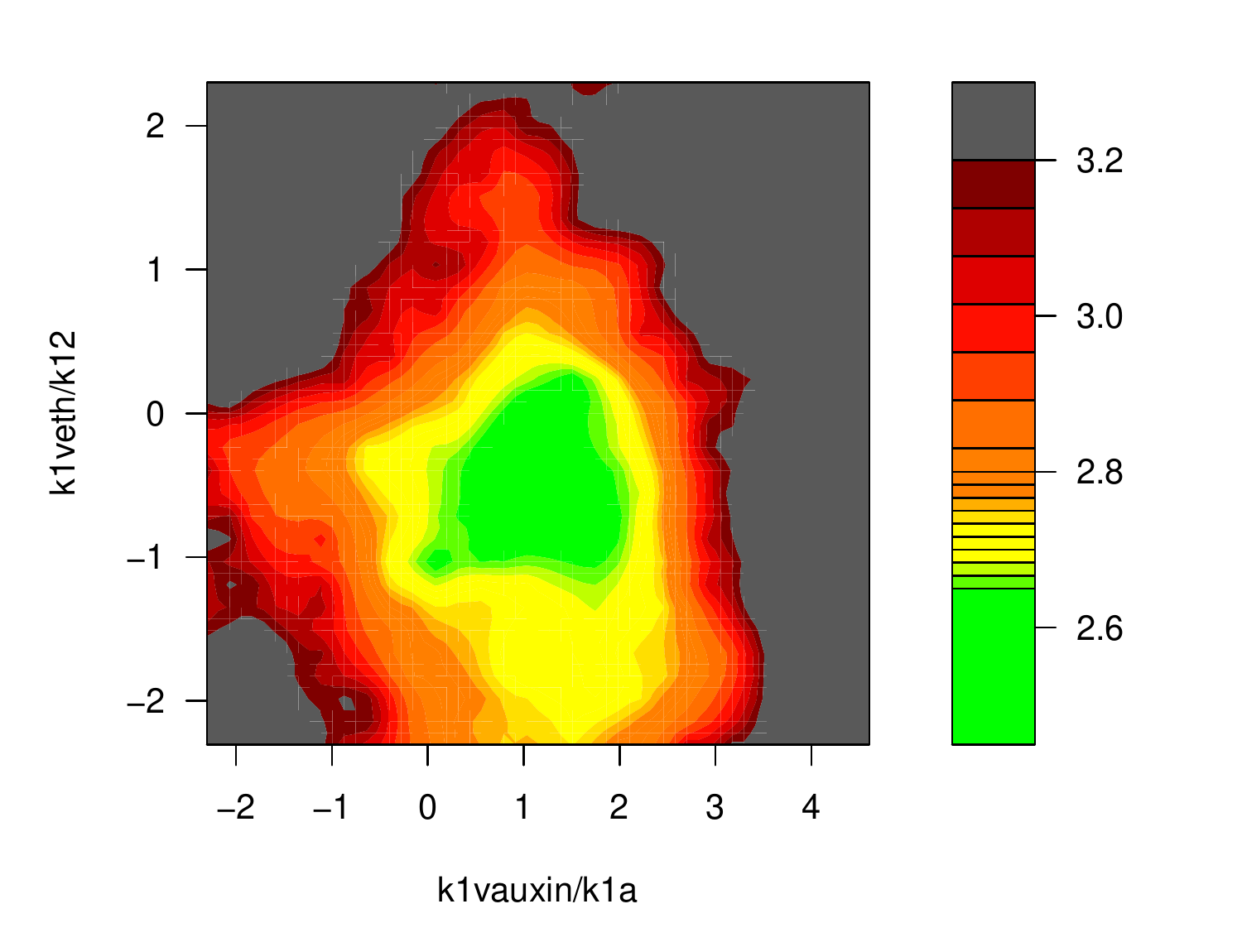} &
\hspace{-0.5cm} \includegraphics[scale=0.421,angle=0,viewport= 42 2 419 310,clip]{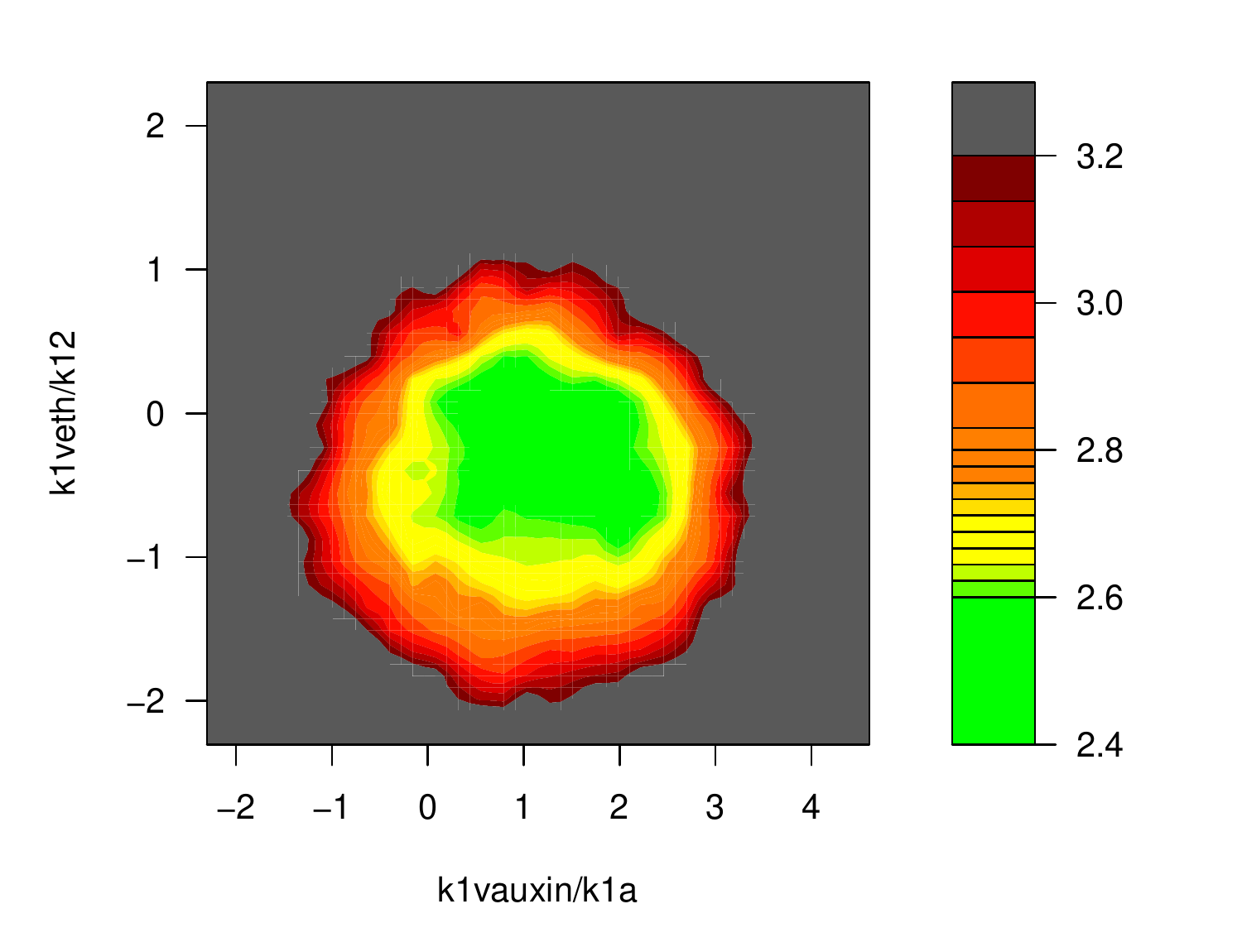} \\
\hspace{-0.65cm} \includegraphics[scale=0.421,angle=0,viewport= 2 2  405 310,clip]{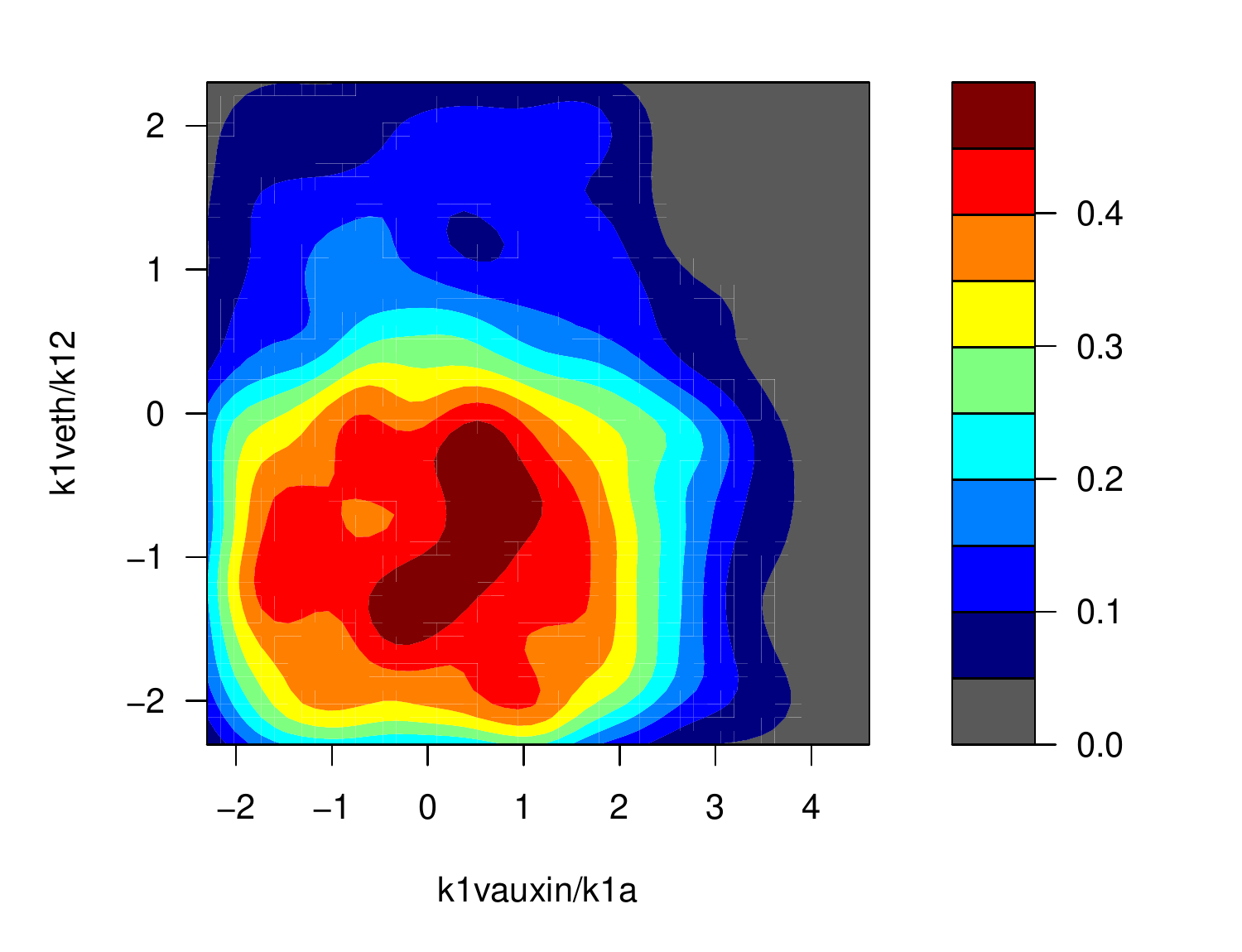} &
\hspace{-0.5cm} \includegraphics[scale=0.421,angle=0,viewport= 42 2 405 310,clip]{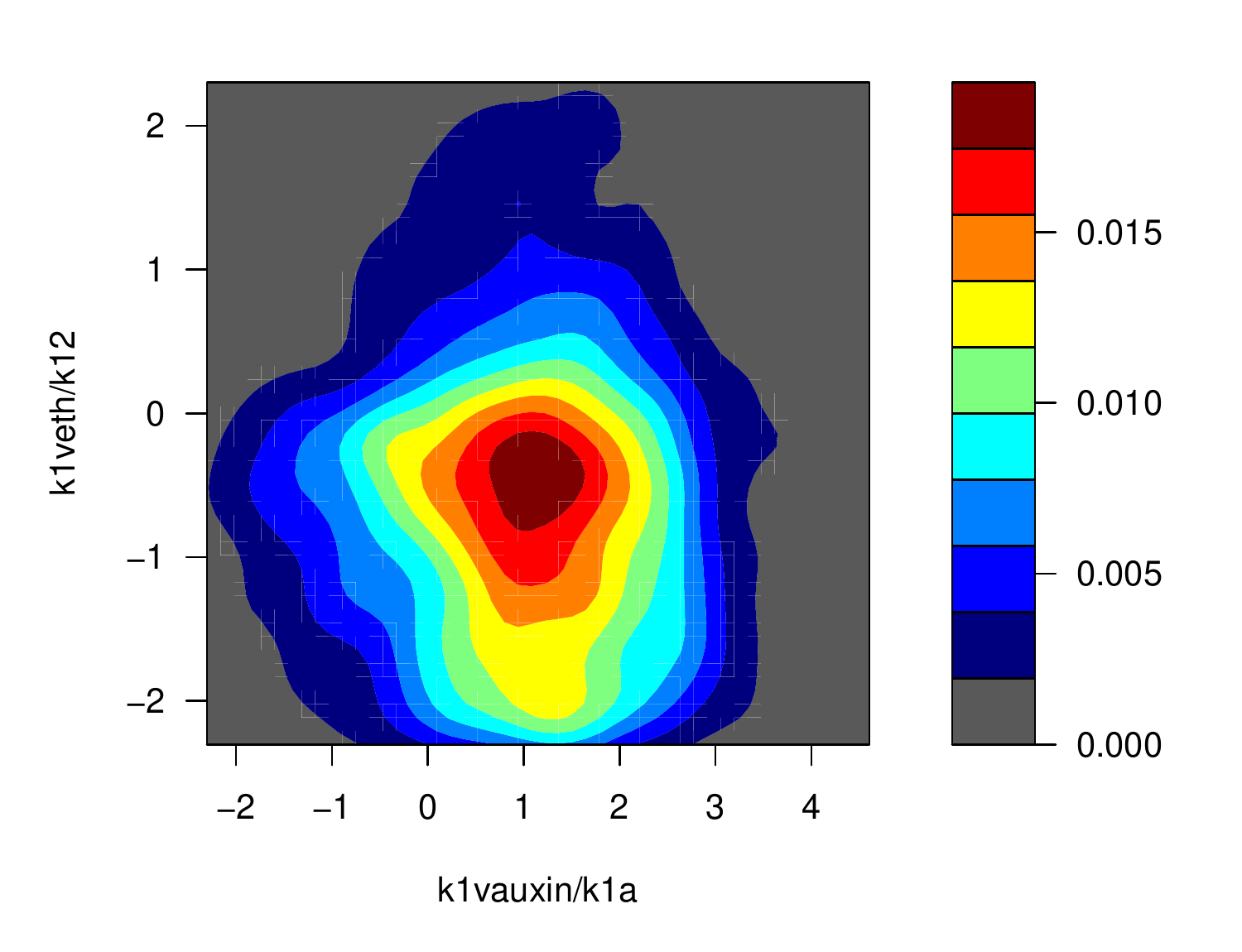} &
\hspace{-0.5cm} \includegraphics[scale=0.421,angle=0,viewport= 42 2 419 310,clip]{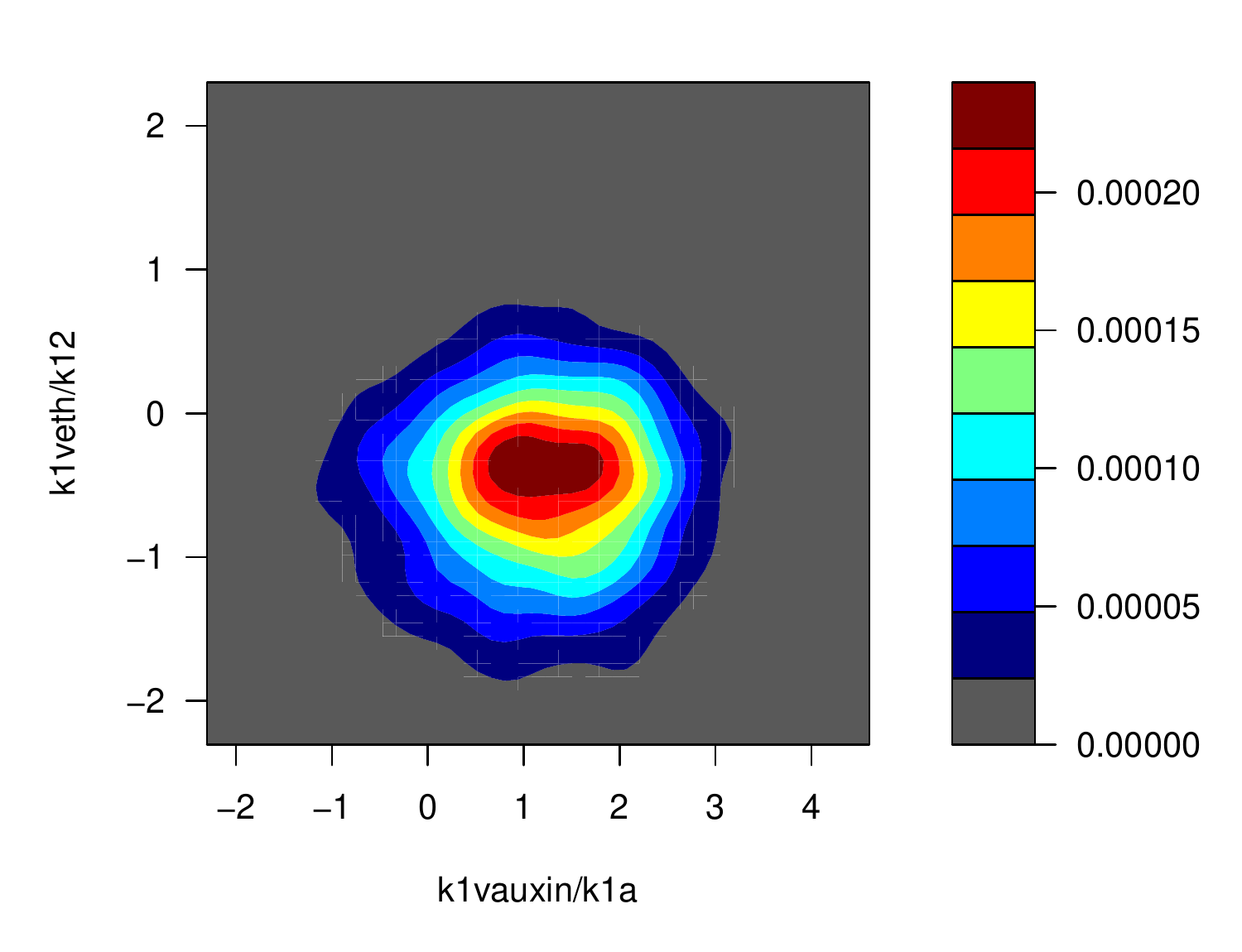}
\end{tabular}
\caption{Two different ways to view the non-implausible input or rate parameter space (on a log scale) after waves 1, 2 and 4 (left, middle and right columns respectively). The top row gives the minimised implausibility $I_P(x')$, where $x' = (k_{1vauxin}/k_{1a}, k_{1veth}/k_{12})$. The red and dark grey regions imply that no matter what values are chosen
for the remaining 21 inputs, the model will still be a poor match to the data for these settings of $k_{1vauxin}/k_{1a}$ and $k_{1veth}/k_{12}$. The bottom row gives the optical depth $\rho(x')$ which shows the thickness or depth of the non-implausible region $\mathcal{X}_k$ in the remaining 21 input dimensions, as a proportion of the depth of the original space $\mathcal{X}_1$.
}\label{fig_suc_waves}
\end{center}
\end{figure*}

Figure~\ref{fig_suc_waves} top left and bottom left panels show two ways of visualising the shape of the non-implausible region $\mathcal{X}_2$ resulting from the wave~1 analysis. The former is the minimised implausibility plot. This is formed by using the emulators to evaluate the implausibility of a large number of points within the 23 dimensional 
$\mathcal{X}_1$. These implausibilities are then projected down to two dimensions (the input parameters $k_{1vauxin}/k_{1a}$ and $k_{1veth}/k_{12}$ in this case) by minimising the implausibility over the remaining 21 dimensions. If we partition $x$ into $(x',x'')$ where $x'$ is the two dimensional vector representing the inputs we wish to project onto and $x''$ is the remaining 21 inputs, then the minimised implausibility is defined as:
\be
I_P(x') \;=\; \min_{x''} I_M(x',x'')
\ee
The plot has the following interpretation: the red/dark grey regions correspond to high implausibility and imply that no matter what values we choose for the remaining 21 inputs, the Arabidopsis model will not give good matches to the data in these regions of ($k_{1vauxin}/k_{1a}$, $k_{1veth}/k_{12}$) space. The green/yellow regions imply that somewhere within the 21-dimensional space there are low implausibility points with these values of $k_{1vauxin}/k_{1a}$ and $k_{1veth}/k_{12}$. We are therefore looking at the 
silhouette of $\mathcal{X}_2$ for various different cutoffs represented as colours~\cite{galf_stat_sci}. The green and yellow regions will be investigated further in subsequent waves. 
 
The bottom left panel of figure~\ref{fig_suc_waves} shows an optical depth plot again for 
the inputs $k_{1vauxin}/k_{1a}$ and $k_{1veth}/k_{12}$. This gives 
the 21 dimensional thickness or depth of $\mathcal{X}_2$ as a proportion of total depth, for each point $x'$ in the 2-dimensional 
($k_{1vauxin}/k_{1a}$, $k_{1veth}/k_{12}$) space. It is defined as
\be
\rho(x') \;=\; \frac{ {\rm V}_{21}\{x\in \mathcal{X}_2\;|\;  x' \ \text{fixed} \} }{ {\rm V}_{21}\{x\in \mathcal{X}_1 \;|\;    x' \ \text{fixed} \}  } ,
\ee
 where ${\rm V}_{21}\{.\}$ denotes the 21-dimensional volume of the remaining space. $\rho(x')$ can therefore show where large or small
amounts of non-implausible points can be found, conditioned on $x'$, providing further insight into the structure of $\mathcal{X}_2$. Both $I_P(x')$ and $\rho(x')$  are generalisable to higher dimensions if necessary, and various computational shortcuts in the emulator calculations can be exploited (see \cite{Vernon10_CS, Vernon10_CS_rej, galf_stat_sci, vernon_astro} for details). 

We then proceeded with a total of 4 waves of emulation and history matching. Summaries of the waves' properties in terms of outputs emulated, numbers of active inputs used, and cutoffs and implausibility measures employed can be found in table~\ref{tab_waves}. The final column gives the proportion of non-implausible space remaining in terms of the original input space, after each wave. At each wave emulator diagnostics are performed by evaluating another 200 model runs over the current non-implausible space, and checking that the 
new emulators predict these 200 runs with appropriate accuracy (see \cite{Tony_EmDiag} for more details on emulator diagnostics). 
\begin{table}
\begin{center}
\begin{tabular}{|c|c|c|ccc|c|}
\hline
Wave & Outputs & Active &  \multicolumn{3}{c|}{Cutoffs $c$ used}  &    Prop. Space\\
& Emul. & Inputs & $I_M$ & $I_{2M}$ & $I_{3M}$  & Non-imp. \\
\hline
 1  & 13 & 7-13  & -    &  3.25  & 3   &     $ 2.06  \times 10^{-1} $  \\
 2  &18  & 6-15 & -    & 3.1  &  2.8  &   $ 4.83  \times 10^{-3} $   \\
 3  & 18 & 6-16   &   5 &  2.9 & 2.7   &   $ 4.34  \times 10^{-4} $  \\
 4  & 18 & 11-19 & 3.2& 2.8 &2.65    &   $ 2.69  \times 10^{-5} $  \\
 5  &  -   &  -   & 3.2 &   -   &  -      &   $ 1.21  \times 10^{-6} $   \\
\hline  
\end{tabular}
\end{center}
\vspace{0.2cm}
\caption{Summary of the 4 waves of emulation. Col. 2: no. of outputs emulated, Col. 3: the no. of active inputs used; Col. 4-6: the implausibility thresholds;
Col. 7: the proportion of the parameter space deemed non-implausible. The 5th wave was performed but not emulated. } \label{tab_waves}
\vspace{-0.3cm}
\end{table}

Figure~\ref{fig_suc_waves} middle and right columns, show the minimised implausibility and optical depth plots after wave 2 and wave 4 respectively, again for the inputs $k_{1vauxin}/k_{1a}$ and $k_{1veth}/k_{12}$, and highlight the progression of the history match and the sequential reduction of the non-implausible space. \red{The minimised implausibility plots also show the sensitivity of the size and location of the non-implausible region to the choice of cutoff motivated by Pukelsheim's rule, and given in table~\ref{tab_waves}}.
Note that in the optical depth plot after wave 4 (bottom right panel), the depth of the non-implausible region is now very small. Even if we were to set the inputs $k_{1vauxin}/k_{1a}$ and $k_{1veth}/k_{12}$ to values corresponding to the largest depth (given by the dark red region), the chances of finding a non-implausible point by randomly choosing the other inputs is approximately $2.3 \times 10^{-4}$, highlighting the difficulty of manual or ad hoc searches of the input space. 

The history matching process is terminated with the evaluation of a wave 5 set of uniformly drawn acceptable runs. 
As shown in table~\ref{tab_waves}, the non-implausible space $\mathcal{X}$ was now $1.21\times 10^{-6}$ smaller than that of the original $\mathcal{X}_1$: a small target, which would require on average a total of 830000 runs chosen at random to obtain 1 single acceptable run, requiring approximately 230 hours of processor time. In contrast, our history matching approach generated hundreds of acceptable runs using only 10000 model evaluations, requiring approximately 2.7 hours of processor time. For a more expensive model in terms of evaluation time, such efficiency gains would be even more dramatic~\cite{Vernon10_CS,Vernon10_CS_rej,galf_stat_sci}.

We now go on to describe the results of the parameter search and discuss their implications.

\begin{figure*}
\vspace{0.cm}
\begin{center}
\hspace{0.cm} \includegraphics[scale=0.53,angle=0]{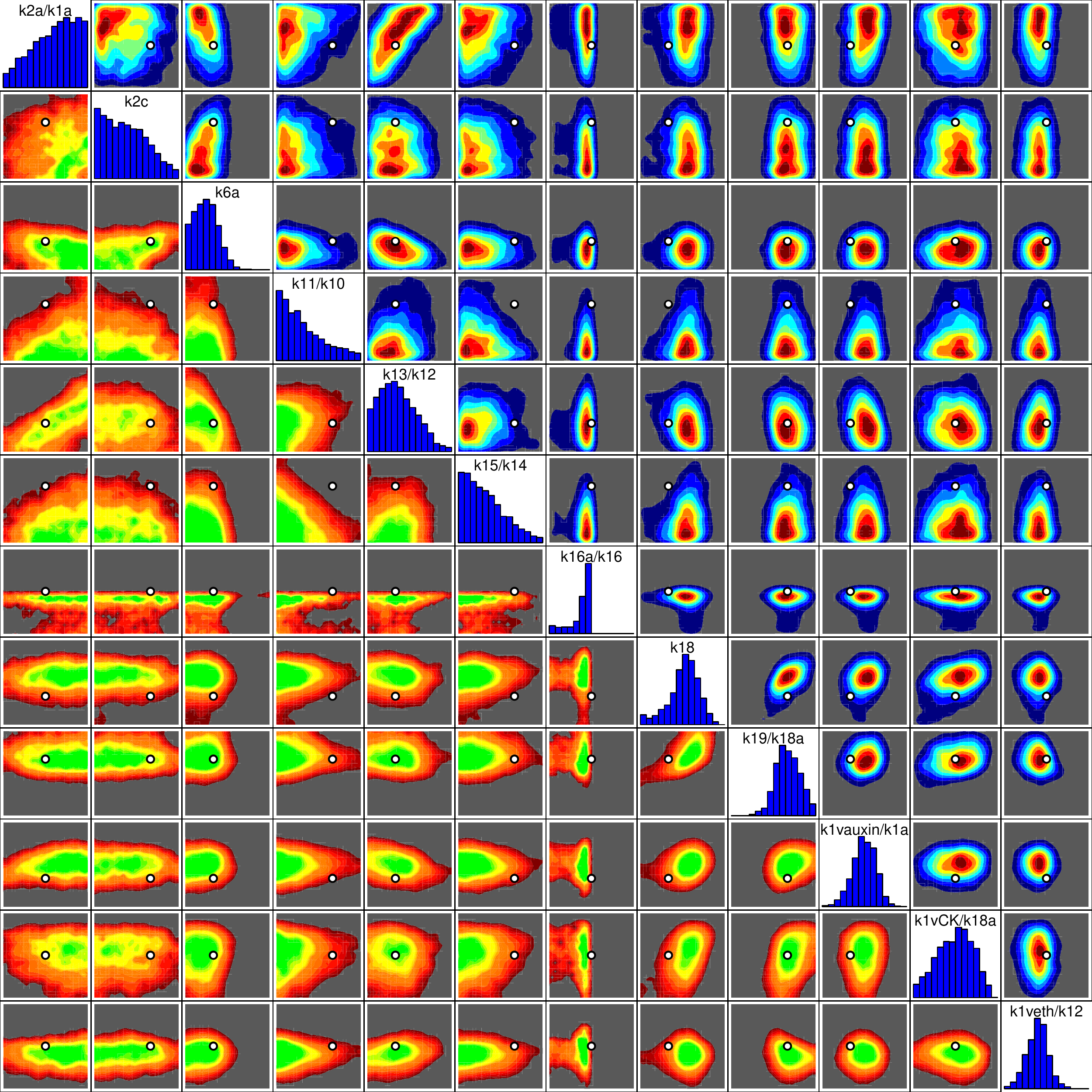}
\caption{The wave 5 minimised implausibility (below diagonal) and optical depth (above diagonal) 
plots for the 12 most 
informed input rate parameters, as labelled along the diagonal. Note that the input rate parameters are on a log scale as given by equation~(\ref{eq_loginputs}) with ranges consistent with table~\ref{tab_input_ranges}. The input location of the previous best run as described in \cite{Liu10_crosstalk} is shown as the single white point. Along the main diagonal, 1-dimensional 
optical depth plots are given.
}\label{fig_mimp_opden}
\end{center}
\end{figure*}

\begin{figure*}
\vspace{0.cm}
\begin{center}
\hspace{0.cm} \includegraphics[scale=0.78,angle=0]{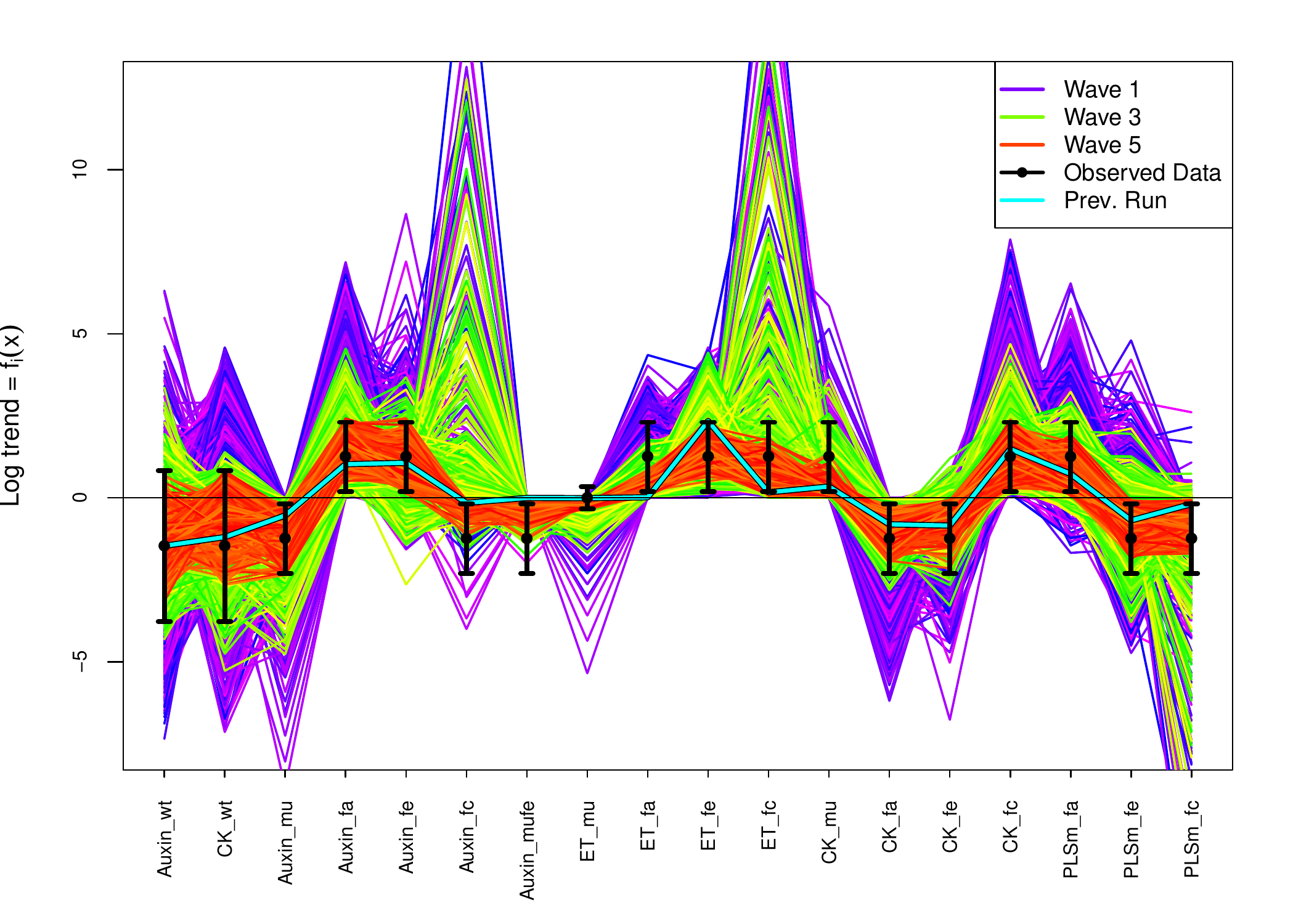}
\caption{The waves 1, 3 and 5 runs as the purple, green and red lines respectively, 
for the 18 model outputs of interest. 
The targets for the history match given by the intervals $z_i \pm 3 \sigma_i$, are shown as the black error bars, and the previous 
best run found by \cite{Liu10_crosstalk} is shown as the light blue line. The horizontal black line at zero represents no trend. 
We see that the history match has proceeded as expected, with the runs from subsequent waves getting closer and closer to the target data, resulting in a large number of acceptable wave 5 runs in red, that possess a better match quality than the previous best run.}\label{fig_outputs_1d}
\end{center}
\end{figure*}

\subsection*{Discussion of the results of the parameter search}

Figure~\ref{fig_mimp_opden} shows the wave 5 minimised implausibility (below diagonal) and optical depth (above diagonal) 
plots for the 12 most 
informed input rate parameters, as labelled along the diagonal. For example, the top right panel gives the optical depth plot 
with $k_{1veth}/k_{12} $ on the x-axis and
$k_{2a}/k_{1a}$ on the y-axis, while the bottom left plot gives the corresponding minimised implausibility plot with the x- and y-axis swapped. 
The input location in parameter space of the previous best run as described in \cite{Liu10_crosstalk} is shown as the single white point in all panels: this corresponds to the single light blue run in figure~\ref{fig_outputs_1d_w1}.
Along the main diagonal, 1-dimensional 
optical depth plots are given, showing that we have learnt most about 
inputs $k_{16a}/k_{16}$, $k_{18}$, $k_{19}/k_{18a}$, $k_{1vauxin}/k_{1a}$ and $k_{1veth}/k_{12} $. All inputs that are not shown 
in this plot were either not constrained at all, or only loosely constrained by the observed data. 
Often, a ``pairs plot" such as shown in  figure~\ref{fig_mimp_opden}, can provide much insight into both the structure of the model and
the complex constraints placed upon the input rate parameters by the data. For example, we instantly see that input k16a/k16 is highly constrained and must lie close to a value of 1/0.3, which we can see is the value that precisely balances the first two terms on the right hand side of the differential equation for $dX/dt$ (given in table~\ref{tab_difeq}), when $CTR1^*$ obtains its maximum value of 0.3.
The $k_{13}/k_{12}$ vs $k_{2a}/k_{1a}$ panel (top row, fifth along from the left) shows a linear relationship (on a log scale) between $k_{13}/k_{12}$ and $k_{2a}/k_{1a}$, in that high values of $k_{13}/k_{12}$ require high values of $k_{2a}/k_{1a}$ to compensate them, and vice versa. The input $k_{6a}$, for which a large range was explored, is constrained to lower values, and has subtle relationships with both the inputs $k_{2a}/k_{1a}$ and $k_{2c}$ (see the panels third from the left in the top two rows). This has important consequences as discussed below. We also see that although the previous best run is close to being an acceptable 
input point, it is not actually contained within the wave 5 non-implausible volume, as can be seen from the $k_{11}/k_{10}$ vs $k_{15}/k_{14}$ plot. This implies that we now have a large number of wave 5 runs that are superior fits to the data than were previously found.
\blue{The minimised implausibility plots of figure~\ref{fig_mimp_opden} also provide, via the implausibility cutoff, insight into the sensitivity of the size and location of the non-implausible region to the original specification of the size of the trend intervals, given by $\sigma_i$ in equation~(\ref{eq_sigma_comb_md_obs}). For example the red regions may be ruled out, were these intervals judged to be moderately less conservative.}

Figure~\ref{fig_outputs_1d} shows the waves 1, 3 and 5 runs as the purple, green and red lines respectively, 
for the 18 model outputs of interest. 
The targets for the history match given by the intervals $z_i \pm 3 \sigma_i$, are shown as the black error bars, and the previous 
best run found by \cite{Liu10_crosstalk} is again shown as the light blue line. Note that the first two error bars correspond to the extra two ouputs of Auxin and CK wildtype with no feeding, while the remaining 16 are the trend data from table~\ref{tab_trends_mod}. The horizontal black line at zero represents no trend. 
We see that the history match has proceeded as expected, with the runs from subsequent waves getting closer and closer to the target data. 
In wave 1, none of the runs simultaneously passed through all the targets, which we now know is due to $\mathcal{X}$ being so small ($1.21\times 10^{-6}$), however we now have hundreds of acceptable runs from within $\mathcal{X}$ shown here as the wave 5 red runs, all of which are a 
better match than the previous best run, and we can quickly generate more. 

Figure~\ref{fig_outputs_1d} also informs as to the class of possible observed data sets that the model could have matched, and hence gives insight into the model's flexibility.
We see that 6 out of the 16 trend outputs could have predicted either positive or negative (or zero) trends, and hence could possibly have fitted many different data sets, although further investigation of the joint structure of these outputs would be required to confirm this. For example, if these 6 outputs were found to vary independently, then 
they could be adjusted to fit {\it any} combination of positive, negative (or zero) trends. However the remaining 10 trend outputs are restricted to giving the `correct' trend, and hence seem not to be flexible at all. In general, we may be concerned about an overly flexible model, possessing say a high number of rate parameters, and specifically about claims that it has been validated based purely on a comparison to data, as it would be no surprise when it fits the observed data well, 
and therefore it may not contain much inherent biological structure at all. 
This is clearly not the case for the Arabidopsis model considered here. Only by performing a global parameter search such as described here, can one guard against such issues. 

We can gain further insight into the model's structure by plotting pairs of outputs against each other, for each wave, as is shown in figure~\ref{fig_pairs_outputs}. Here the colour scheme is consistent with figure~\ref{fig_outputs_1d} with the wave 1, 3 and 5 runs as purple, green and red points respectively, the target intervals are now represented as 2D boxes and the previous best run given as the light blue point. The top right 
panel, for example, shows the Auxin output for the {\it pls} mutant strain (Auxin\_mu) on the y-axis and the PLSm output with Auxin feeding (PLSm\_fa) on the x-axis. This suggests that large negative trends for the Auxin\_mu output can only occur when the PLSm\_fa trend is close to zero. 
Similarly, a high 
PLSm\_fa trend implies Auxin\_mu must also be close to zero. These plots also highlight previously unknown model constraints between the outputs e.g. the Auxin\_fa vs CK\_fa panel shows that these two trends satisfy a strict inequality in log space that bisects the target box. Similar strict constraints are seen in the Auxin\_mu vs CK\_mu and Auxin\_fa vs PLSm\_fa panels. We now go onto discuss in more
detail the implications for gene functions of the parameter search results.

\begin{figure*}
\vspace{0.cm}
\begin{center}
\hspace{0.cm} \includegraphics[scale=1.0,angle=0,viewport=-1 -1 350 350,clip]{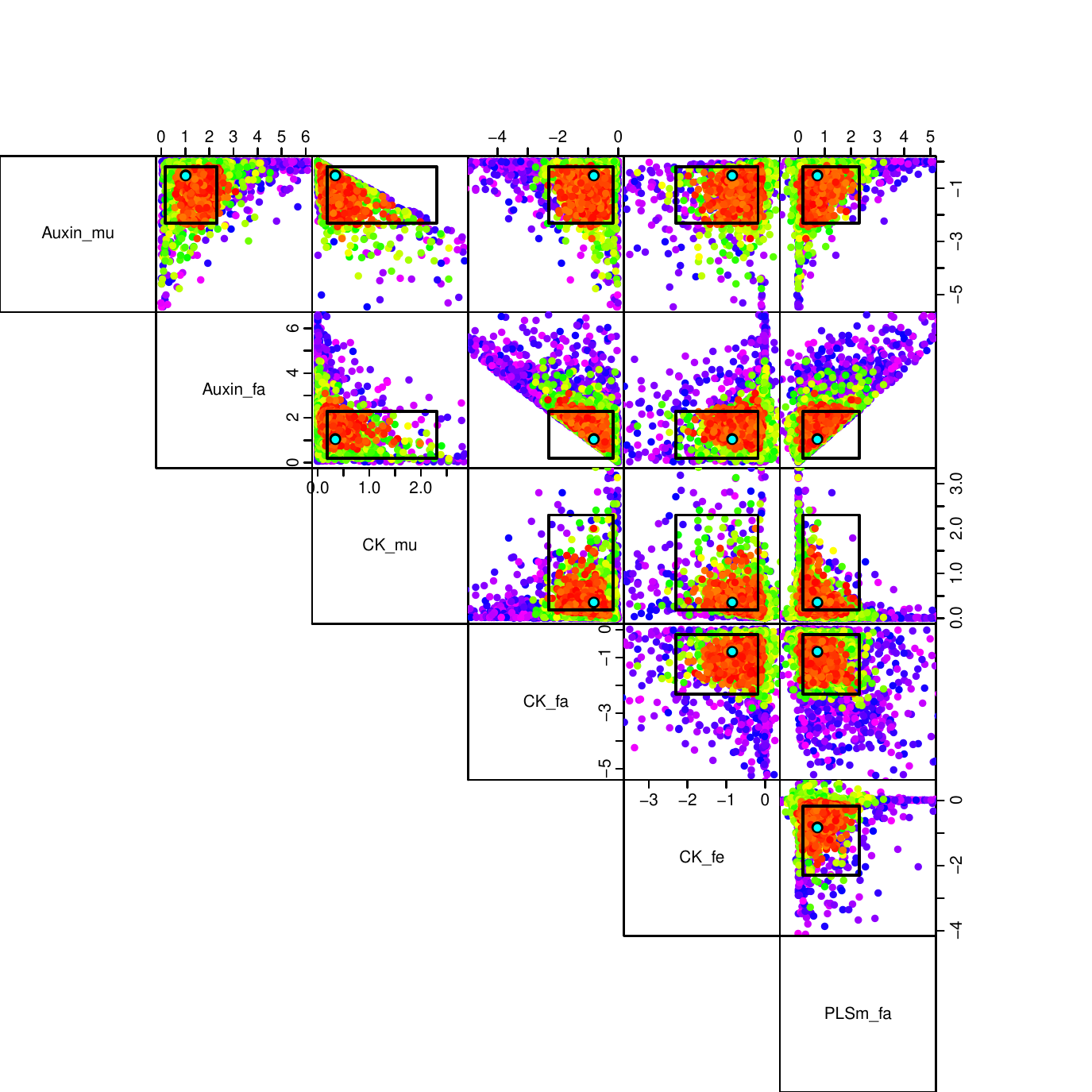}
\caption{We can gain further insight into the model's structure by plotting pairs of outputs against each other, for each wave. Here the colour scheme is consistent with figure~\ref{fig_outputs_1d} with the wave 1, 3 and 5 runs as purple, green and red points respectively and the outputs labelled along the diagonal. The target intervals are now represented as 2D boxes and the previous best run given as the light blue point. 
We can hence see several hard constraints between the model's outputs, for example between Auxin\_fa and CK\_fa.
}\label{fig_pairs_outputs}
\end{center}
\end{figure*}

\section*{Discussion}
\subsection*{Evaluation of gene functions using Bayesian emulation methodology}

In the previous sections, we have shown that Bayesian emulation and history matching methodology allows extensive exploration of the input rate parameter space, giving multiple insights into the model's structure, constraints placed upon it by the observed data and on the corresponding biological consequences. 

Here we further demonstrate that this methodology can be used to evaluate regulatory relationships and gene functions in hormonal signalling systems, by examining both the above results and the results obtained from a second history match of an alternative model. 
 
The $k_{6a}$ rate parameter describes the regulatory strength of ethylene as applied to the PLS transcriptional rate. It features in the first term on the right hand side of the $d[PLSm]/dt$ equation in table~\ref{tab_difeq}, and in the limit $k_{6a}\rightarrow \infty$ we have that
\be
\frac{k_6  [Ra^*]} { 1 + \frac{[ET]}{k_{6a}}}  \quad \longrightarrow \quad   k_6  [Ra^*] 
\ee
Therefore increasing $k_{6a}$ decreases this regulatory strength. Thus, low values of $k_{6a}$ indicate that a regulatory relationship of ethylene inhibiting PLSm production is required. The optical depth and minimised implausibility plots corresponding to $k_{6a}$ in Figure 9 show that high values of $k_{6a}$ are ruled out. Our analysis suggests that no acceptable parameter combinations with large $k_{6a}$ can be found that are consistent with the target data, and hence our results strongly support the assertion that the inhibition of PLSm production by ethylene is required for predicting known experimental trends, conditional on the remaining specifications made in the analysis.

The k2c parameter describes the very important question of whether the {\it PLS} gene has a function in auxin biosynthesis. 
Examining the third term on the right hand side of the $d[Auxin]/dt$ equation in table~\ref{tab_difeq}, we see that as $k_{2c}\rightarrow 0$ we have that
\be \label{eq_k2c}
\frac{k_{2a} [ET] }{ 1 + \frac{[CK]}{k_{2b}}}  \frac{[PLSp]} { k_{2c} + [PLSp]}  \longrightarrow \frac{k_{2a} [ET] }{ 1 + \frac{[CK]}{k_{2b}}}
\ee
Therefore the $k_{2c}=0$ case implies that the {\it PLS} gene has no direct function in auxin biosynthesis, where the $k_{2c}>0$ case would imply that it does. 
However, for several of the outputs considered, $[PLSp]$ can also tend to zero, implying that the limit given in equation~(\ref{eq_k2c}) is not uniquely defined, and that 
the original model is not continuous at $k_{2c}=0$. Hence, to answer questions regarding the role of the {\it PLS} gene in auxin biosynthesis we cannot simply examine low values of $k_{2c}$. As the $k_{2c}=0$ case effectively defines a distinct model, we perform a new 5 wave history match to find any acceptable matches to the observed data, following the same methodology as described above. The results of the new history match are given in figure~\ref{fig_outputs_1d_k2c}, and notably we  
again found several acceptable wave 5 runs shown as the red lines, that are in agreement with the observed trends. The acceptable runs were found in a smaller region than previously, with a volume of $\mathcal{X}$ approximately $2.4 \times 10^{-8}$ of that of $\mathcal{X}_1$.

\begin{figure*}
\vspace{0.cm}
\begin{center}
\hspace{0.cm} \includegraphics[scale=0.78,angle=0]{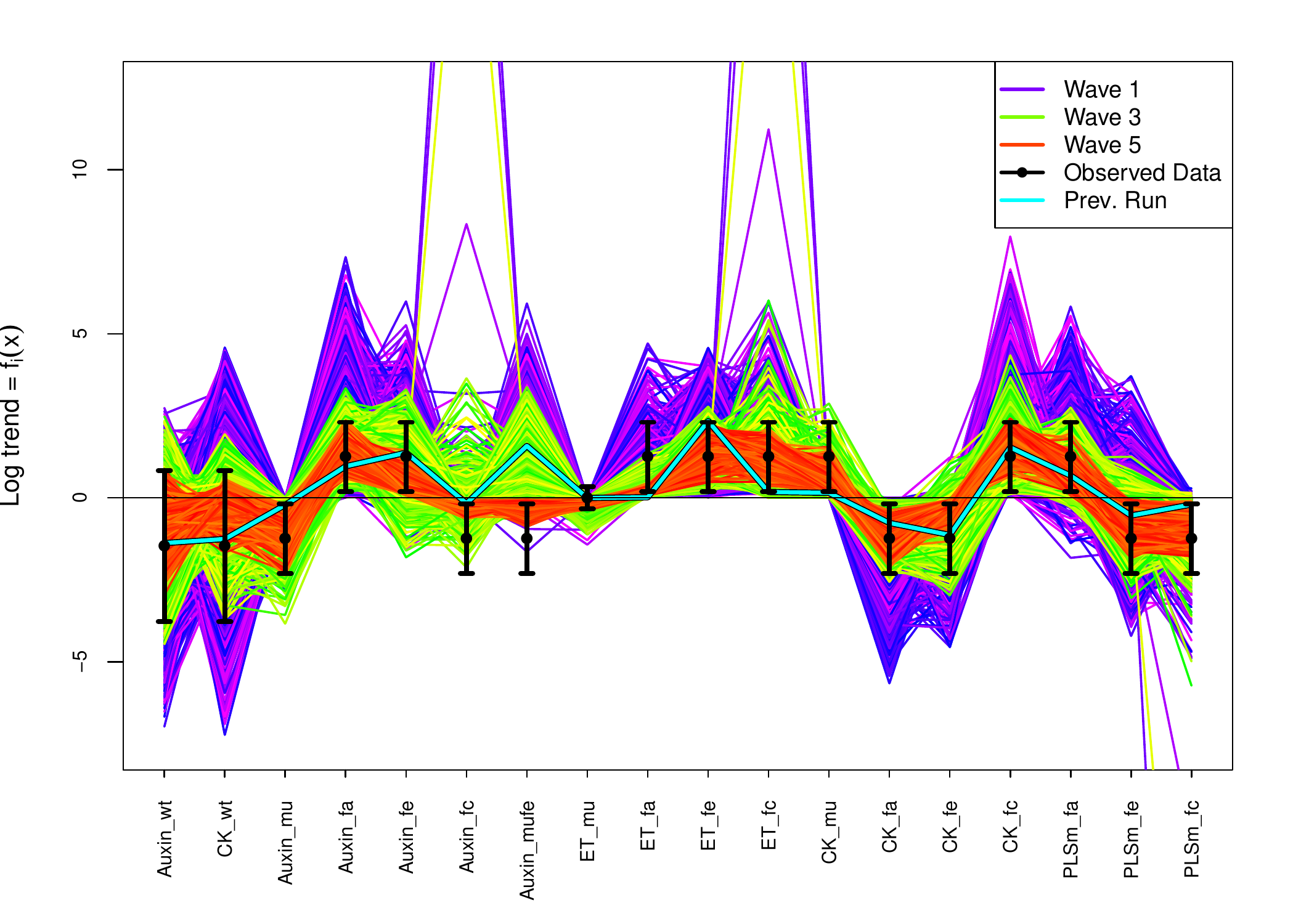}
\caption{The results from the history match of the $k_{2c}=0$ model, showing the waves 1, 3 and 5 runs as the purple, green and red lines respectively, 
for the 18 model outputs of interest. We see that a large number of acceptable wave 5 runs have been found that match the target data (the black error bars) and
hence that the reduced $k_{2c}=0$ model is still consistent with the specified observational data, within a small region of the input space. Comparison with figure~\ref{fig_outputs_1d} shows the main differences between the two models, most noticeably in the Auxin\_mufe output (7th error bar from the left) for which the vast majority of the input space returns an incorrect positive trend for the $k_{2c}=0$ model, as can be seen by the large number of wave 1 and 3 lines above zero, with only a small and hence hard to find region returning the correct negative trend. The $k_{2c}>0$ model conversely, always returns the correct negative trend.}\label{fig_outputs_1d_k2c}
\end{center}
\end{figure*}

Comparing the results of the $k_{2c}=0$ case (figure~\ref{fig_outputs_1d_k2c}) with the results of the $k_{2c}>0$ case (figure~\ref{fig_outputs_1d}) we can 
immediately see some important differences between the two models. 
For the Auxin\_mufe output (7th error bar from the left), the $k_{2c}>0$ model always returns the correct negative trend. In contrast the $k_{2c}=0$ 
model returned the incorrect positive trend for the vast majority of the wave 1, 2 and 3 runs, implying that there is only a very small region of input space that 
returns the correct negative trend, a region located by the history match analysis and explored by the wave 5 runs. Without such an analysis it would be easy to 
incorrectly conclude that the $k_{2c}=0$ model is \red{inconsistent with the data}. 
This demonstrates perhaps the {\bf most important difficulty} in exploring high dimensional models: there may be one (or more) extremely 
small regions of input space of scientific interest, and conventional optimisation techniques may easily get stuck in local minima far away from these regions. Our 
Bayesian history matching approach however is specifically designed to combat such difficulties by carefully exploring the space using efficient emulator based global search methods, as we have demonstrated here. 

After considering that the {\it PLS} gene is required for the response of ethylene downstream based on experimental observations (mathematically this is equivalent to the response of ethylene downstream, X, remaining constant for the {\it pls} mutant ($k_6$=0) fed with ethylene), previous research~\cite{Liu10_crosstalk} deduced that the {\it PLS} gene does indeed have a function in auxin biosynthesis. However, the history match of the $k_{2c} = 0$ model (figure~\ref{fig_outputs_1d_k2c}) suggests that, 
given the specification of the trends and their relevant uncertainties, the $k_{2c} = 0$ model is consistent with observed data, 
and hence it may not be essential for the {\it PLS} gene to play a role in auxin biosynthesis.

However, examining the differences between the two models reveals some interesting results.
Figure~\ref{fig_outputs_x4} summarises the history match results of both the $k_{2c}>0$ and $k_{2c}=0$ models. 
\begin{figure*}
\vspace{0.cm}
\begin{center}
\hspace{0.cm} \includegraphics[scale=0.78,angle=0]{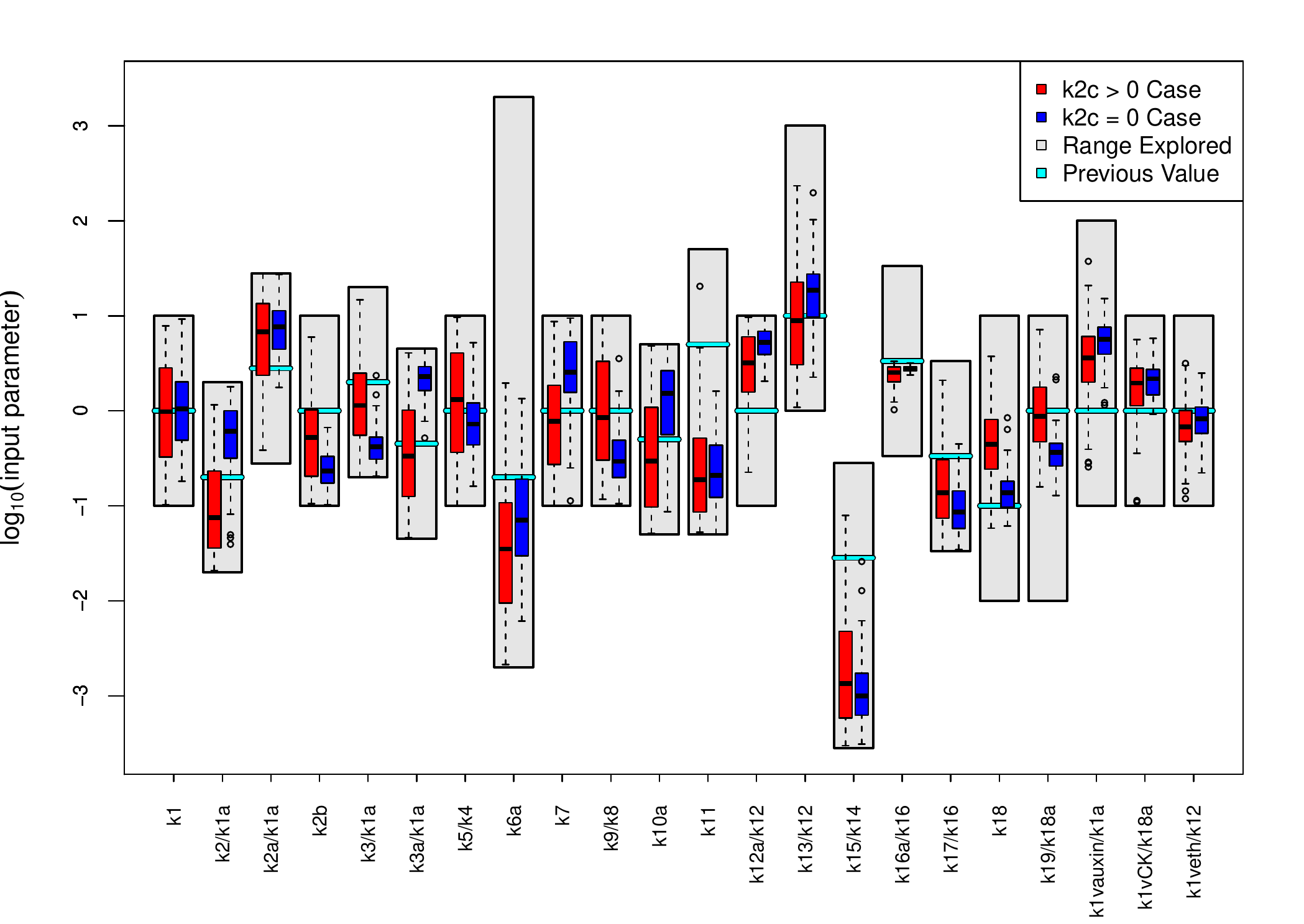}
\caption{A comparison of the spread of input parameter locations of the acceptable runs found for the $k_{2c}>0$ model (red boxplots) and the $k_{2c}=0$ model (blue boxplots) 
in terms of individual inputs as labelled along the x-axis. Note that the two sets of acceptable runs being compared correspond to the 
red lines in figures~\ref{fig_outputs_1d} and \ref{fig_outputs_1d_k2c} respectively. The y-axis is on a $\log_{10}$ scale, and the grey rectangles show the initial ranges that define the original search region $\mathcal{X}_1$ as given in table~\ref{tab_input_ranges}. The light blue horizontal lines show the input parameter values of the previous best run as found by~\cite{Liu10_crosstalk}.
We can see that some inputs, for example, $k_1$, are either unconstrained by the 
observed data, or possibly that changes in these inputs can be compensated by appropriate changes in other inputs. Some inputs such as $k_{16a}/k_{16}$ are highly constrained, while 
others such as $k_2/k_{1a}$, $k_{2b}$, $k_3/k_{1a}$ and $k_{3a}/k_{1a}$ show clear differences between the $k_{2c}>0$ and $k_{2c}=0$ models, with the latter model preferring 
higher values of $k_2/k_{1a}$ and $k_{3a}/k_{1a}$ and lower values of $k_{2b}$ and $k_3/k_{1a}$. 
}\label{fig_outputs_x4}
\end{center}
\end{figure*}
It shows a comparison of the spread of input parameter locations of the acceptable runs found for the $k_{2c}>0$ model (red box plots) and the $k_{2c}=0$ model (blue box plots) 
in terms of individual input rate parameters as labelled along the x-axis. Note that the two sets of acceptable runs being compared correspond to the 
red lines in figures~\ref{fig_outputs_1d} and \ref{fig_outputs_1d_k2c} respectively. The y-axis is on a $\log_{10}$ scale, and the grey rectangles show the initial ranges that define the original search region $\mathcal{X}_1$ as given in table~\ref{tab_input_ranges}. 
The light blue horizontal lines show the input parameter values of the previous best run as found by~\cite{Liu10_crosstalk}. The main differences between the two models' acceptable runs are exhibited by the following parameters or ratios of parameters: $k_2/k_{1a}$, $k_{2b}$, $k_3/k_{1a}$, $k_{3a}/k_{1a}$, $k_{12a}/k_{12}$, $k_{15}/k_{14}$, $k_{18}$, $k_{19}/k_{18}$. To the best of our knowledge, the biological significance of many of these differences cannot be judged using current biological insight. However, two ratios, $k_2/k_{1a}$ and $k_{3a}/k_{1a}$, do reveal some important results. $k_{1a}$ is the maximal rate of transporting auxin from shoot to root; $k_2$ is the background auxin biosynthesis rate; $k_{3a}$ is the rate constant describing the control of ethylene downstream over auxin transport from root to shoot.  First, biologically $k_2/k_{1a}$ must be very small. This is because the background auxin biosynthesis rate, $k_2$,  must be very small and usually biologically negligible, as it represents the non-enzymatic process in auxin biosynthesis. Moreover, auxin transport from shoot to root, whose maximal rate is $k_{1a}$, is an important process, as evidenced experimentally~\cite{Vanneste:2009aa}. Therefore, $k_{1a}$ should be large. However, for the $k_{2c} = 0$ model to match target data, the majority of acceptable runs have relatively large $k_2/k_{1a}$, while for the $k_{2c} > 0$ model much smaller and more realistic values are preferred. Second, biologically $k_{3a}/k_{1a}$ should be small. This is because it is known that auxin more predominantly transports from shoot to root, to form an auxin concentration maximum in the root tip \cite{Friml2002661,Sabatini1999463}. However, for the $k_{2c} = 0$ model to match target data, the set of acceptable runs suggest that relatively large $k_{3a}/k_{1a}$ is required. Therefore, the differences between the two models' parameter ratios highlight that, although we have found acceptable matches for the $k_{2c} = 0$ model, these matches have not been found at biologically realistic parameter values. While we must be cautious about such conclusions that are based on the finite sampling of the non-implausible regions, we have generated hundreds of approximately uniformly sampled acceptable runs from each model that do indeed exhibit the features discussed. 
Therefore our results suggest that biological insight clearly favours the 
model with $k_{2c}>0$, i.e. that the {\it PLS} gene does have a function in auxin biosynthesis. More detailed measurements of the key outputs that restrict $k_2/k_{1a}$ and $k_{3a}/k_{1a}$ would of course further clarify this issue.

Our results show that Bayesian emulation and history matching methodology can be used to evaluate regulatory relationships and gene functions in hormonal signalling systems. To further improve the accuracy of the results of this methodology, the following aspects should be considered. First, experimental data should be more quantitatively measured, to define more accurate trends. The example trends we have used in this work, as summarised in table~\ref{tab_trends_mod} and the associated discussion, are mainly formulated based on qualitative or semi-quantitative experimental data, combined with scientific judgement. Second, model development should include more components, to better describe the experimental systems. Third, 
Bayesian emulation methodology should be used to study the effects of additional experiments, such as the response of ethylene downstream when feeding ethylene, {\it etr1} mutant and {\it etr1-pls} double mutant, on the evaluation of regulatory relationships and gene functions. Fourth, 
Bayesian emulation methodology should also be used to explore the effects of the uncertainty of quantitative trends on the evaluation of regulatory relationships and gene functions, as in most cases trends of biological data are not sufficiently quantitative.

\section*{Conclusions}

We have provided an introduction to the study of complex systems biology models using Bayes linear uncertainty analysis. This represents
a possible solution to the fundamental challenge that faces systems biology in terms of the necessity of global parameter searches 
of high dimensional models. 
Our approach features three main aspects:
\bi
\item A more formal statistical model linking the biological model to reality, which encompasses major sources of uncertainty such as observational errors and model discrepancy.
\item A Bayesian emulator allowing a very fast exploration of model behaviour, applicable to models even with very long evaluation times.
\item A careful history match using implausibility measures that performs an iterative global exploration of the input parameter space using the emulators, to find the region containing {\it all} acceptable matches to the observed data.
\ei

We applied this methodology to two versions of the hormonal crosstalk in Arabidopsis root development model, and in each case identified the small region of input space containing scientifically interesting matches. 
The two models and their biological implications were then compared in a robust manner and used to discuss gene functions.
We found that although some acceptable matches to the specified trends could be found for the $k_{2c}=0$ model, these were only found at parameter 
settings that violated other known biological evidence, whereas the $k_{2c}>0$ model's acceptable matches seemed far more realistic. This implied that 
PLS does indeed play a role in auxin biosynthesis. 
Our results also strongly supported the assertion that the inhibition of PLSm production by ethylene is required for consistency with known experimental trends.

We would stress that searching for all acceptable matches between model output and observed data is vital for several reasons. It avoids the danger of false conclusions being made, based on the analysis of a single run (or a small number of runs) consistent with the data: conclusions that could easily change if an alternative run was found instead, that also matched the data but which provided different biological implications.  
If we want to use the model to make predictions for the results of future biological experiments, all acceptable matches must be found and the corresponding range of predictions examined. A narrow range of predictions from the acceptable runs for a particular proposed future experiment, for example, would imply that it would be a good test of the model as it could possibly rule it out, while a large range implies that this experiment would most likely be informative for the model's rate parameters. Model predictions, using all the acceptable runs, can then be used to design efficient sets of future experiments that are most likely to realise particular 
scientific goals, such as learning about all or subsets of the rate parameters, testing the model or distinguishing between certain biological hypotheses. 
We would assert that this design problem is also a fundamental challenge to the area of systems biology, but leave a detailed exposition to future work~\cite{Ver_design_sysbio_stats,Ver_design_sysbio_bio}.

Since plant root development is regulated by multiple hormones in a coordinated way \cite{DVDY:DVDY23878}, unravelling the regulatory relationships and gene functions for root development is a difficult task that requires the investigation of how biological information is spatiotemporally integrated and communicated \cite{Chaiwanon-J:2016aa}. Modelling hormonal crosstalk as an integrative system is an important aspect for integrating information in plant root development \cite{Liu10_crosstalk,10.3389/fpls.2013.00075,10.3389/fpls.2014.00116,NPH:NPH13421, Moore:2015aa, NPH:NPH13882}. This work demonstrates that a combination of experimental data, a model of hormonal crosstalk in Arabidopsis root development, and Bayesian emulation and history matching methodology is able to evaluate regulatory relationships and gene functions in a hormonal signalling system. In particular, Bayesian emulation and history matching methodology is a \red{useful} method for performing a global parameter search to attempt to find all input parameter settings that achieve an acceptable match.


\begin{backmatter}

\section*{Ethics approval and consent to participate}
Not applicable.

\section*{Consent for publication}
Not applicable.

\section*{Availability of data and material}
All data generated or analysed during this study are included in this published article.

\section*{Competing interests}
The authors declare that they have no competing interests.

\section*{Funding}
This work was initiated as part of an EPSRC seed corn grant, administered by Durham University. In addition, JL and KL gratefully acknowledge the Biotechnology \& Biological Sciences Research Council (BBSRC) for funding in support of this study. JR is in receipt of a BBSRC studentship. IV and MG gratefully acknowledge MRC and EPSRC funding.

\section*{Author's contributions}
IV, JL, MG and KL conceived the idea. IV and JL designed this research and carried out the data analysis. IV and MG developed and performed  the Bayesian emulation uncertainty analysis. All authors established the links between the Bayesian emulation methodology and hormonal crosstalk in plant development. IV, JL, MG and KL wrote the manuscript.

\section*{Acknowledgements}

JL and KL gratefully acknowledge the Biotechnology \& Biological Sciences Research Council (BBSRC) for funding in support of this study. JR is in receipt of a BBSRC studentship. IV and MG gratefully acknowledge MRC and EPSRC funding.


\bibliographystyle{bmc-mathphys} 
\bibliography{master_bib}

\section*{Additional Files}
  \subsection*{Additional file 1 - ``History\_matching\_for\_systems\_biology\_Supplementary\_Material.pdf"}
 This file gives the details of the full emulator structure used in the wave 1 emulators, described in the subsection entitled ``Bayesian emulation of the Arabidopsis model", 
 and gives the dimensions or units of all the rate constants.
  \subsection*{Additional file 2 - ``R\_code\_for\_1D\_Example.R"}
R code to reproduce the 1D example model output, discrepancy, emulation and history matching plots of figures 1, 2, 4 and 5 respectively.

%

\end{backmatter}

\end{document}